\documentclass[leqno]{amsart}

\usepackage[foot]{amsaddr}

\usepackage{color}
\usepackage{geometry}
\usepackage{subfiles}
\usepackage{graphicx}
\usepackage[section]{placeins}
\usepackage{hyperref}
\usepackage{stmaryrd}
\usepackage{amsmath,amssymb,amsfonts,amsthm,textcomp}
\usepackage{mathtools}
\usepackage{tensor}
\usepackage{multicol}
\usepackage{subfig}
\usepackage{enumitem}
\usepackage{tcolorbox}

\graphicspath{
              {./figures/}
             }

\newfont{\tenbfsl}{cmbxti9 scaled 1200}
\newfont{\tenbbb}{msbm10}
\newfont{\svnbbb}{msbm8}

\newcommand{\sqr}[1]{\overset{\scriptscriptstyle\square}{#1}}

\newcommand{\bs}[1]{\boldsymbol{#1}}
\newcommand{\br}[1]{\boldsymbol{\mathrm{{#1}}}}

\newcommand{\cl}[1]{\mathcal{#1}}

\newcommand{\bb}[1]{\mathbb{#1}}

\newcommand{\id}{\bs{1}}

\newcommand{\surp}[1]{\{\!\!\{ {#1} \}\!\!\}}

\newcommand{\fr}[2]{{\textstyle{\frac{{#1}}{{#2}}}}}

\newcommand{\dv}{\,\mathrm{d}v}
\newcommand{\da}{\,\mathrm{d}a}
\newcommand{\ds}{\,\mathrm{d}\sigma}

\newcommand{\prt}{\cl{P}}
\newcommand{\dprt}{\partial\cl{P}}
\newcommand{\ddprt}{\partial^2\cl{P}}
\newcommand{\srf}{\cl{S}}
\newcommand{\dsrf}{\partial\cl{S}}

\newcommand{\edg}{\cl{C}}

\newcommand{\intPt}{\int\limits_{\prt_\tau}}
\newcommand{\intdPt}{\int\limits_{\dprt_\tau}}
\newcommand{\intddPt}{\int\limits_{\ddprt_\tau}}

\newcommand{\intR}{\int\limits_{\cl{R}}}
\newcommand{\intRt}{\int\limits_{\cl{R}_\tau}}

\newcommand{\trans}{\scriptscriptstyle\mskip-1mu\top\mskip-2mu}

\newcommand{\tr}{\mathrm{tr}\mskip2mu}
\newcommand{\sym}{\mathrm{sym}\mskip2mu}

\newcommand{\Grad}{\mathrm{grad}\mskip2mu}
\newcommand{\Div}{\mathrm{div}\mskip2mu}

\newcommand{\Grads}{\Grad_{\mskip-2mu\scriptscriptstyle\cl{S}}}
\newcommand{\Divs}{\Div_{\mskip-6mu\scriptscriptstyle\cl{S}}}
\newcommand{\triangles}{\triangle_{\mskip-2mu\scriptscriptstyle\cl{S}}}

\theoremstyle{remark}
\newtheorem{rmk}{Remark}

\theoremstyle{definition}

\newcommand{\twovdots}{\mskip+2mu\colon\mskip-2mu}
\makeatletter
\def\threevdots{\mskip+4mu\vbox{\baselineskip2.25\p@ \lineskiplimit\z@
  \kern4.9\p@\hbox{.}\hbox{.}\hbox{.}}\mskip+3.8mu}
\makeatother

\newcommand{\Prj}[1]{\br{P}_{\scriptscriptstyle{\mskip-6mu{#1}}}}

\newcommand{\bsts}{\bs{t}_{\scriptscriptstyle\cl{S}}}
\newcommand{\bshds}{\bs{h}_{\scriptscriptstyle\partial\cl{S}}}

\newcommand{\xis}{\xi_{\scriptscriptstyle\cl{S}}}
\newcommand{\tauds}{\tau_{\scriptscriptstyle\partial\cl{S}}}

\newcommand{\brFP}{\br{F}_{\!\scriptscriptstyle\prt}}
\newcommand{\dbrFP}{\dot{\br{F}}_{\!\scriptscriptstyle\prt}}

\newcommand{\brLP}{\br{L}_{\!\scriptscriptstyle\prt}}
\newcommand{\JP}{J_{\scriptscriptstyle\prt}}
\newcommand{\dJP}{\dot{J}_{\scriptscriptstyle\prt}}

\newcommand{\brFdP}{\br{F}_{\!\scriptscriptstyle\dprt}}
\newcommand{\dbrFdP}{\mathring{\br{F}}_{\!\scriptscriptstyle\dprt}}

\newcommand{\brLdP}{\br{L}_{\!\scriptscriptstyle\dprt}}
\newcommand{\JdP}{J_{\scriptscriptstyle\dprt}}
\newcommand{\dJdP}{\mathring{J}_{\scriptscriptstyle\dprt}}

\newcommand{\pP}{p_{\scriptscriptstyle\cl{P}}}
\newcommand{\pPtot}{p_{\scriptscriptstyle\cl{P}}^{\mathrm{tot}}}
\newcommand{\pPcap}{p_{\scriptscriptstyle\cl{P}}^{\mathrm{cap}}}
\newcommand{\pPmech}{p_{\scriptscriptstyle\cl{P}}^{\mathrm{mech}}}

\newcommand{\pdPtot}{p_{\scriptscriptstyle\partial\cl{P}}^{\mathrm{tot}}}
\newcommand{\pdPcap}{p_{\scriptscriptstyle\partial\cl{P}}^{\mathrm{cap}}}
\newcommand{\pdPtherm}{p_{\scriptscriptstyle\partial\cl{P}}^{\mathrm{therm}}}
\newcommand{\pdPmech}{p_{\scriptscriptstyle\partial\cl{P}}^{\mathrm{mech}}}

\newcommand{\bmuP}{\bar{\mu}_{\scriptscriptstyle\cl{P}}}
\newcommand{\bmudP}{\bar{\mu}_{\scriptscriptstyle\partial\cl{P}}}

\newcommand{\barkappadP}{\bar{\kappa}_{\scriptscriptstyle\partial\cl{P}}}

\newcommand{\phiP}{\phi_{\scriptscriptstyle\cl{P}}}
\newcommand{\phidP}{\phi_{\scriptscriptstyle\partial\cl{P}}}
\newcommand{\dphiP}{\dot{\phi}_{\scriptscriptstyle\cl{P}}}
\newcommand{\dphidP}{\mathring{\phi}_{\scriptscriptstyle\partial\cl{P}}}
\newcommand{\rhoP}{\varrho_{\scriptscriptstyle\cl{P}}}
\newcommand{\rhodP}{\varrho_{\scriptscriptstyle\partial\cl{P}}}
\newcommand{\drhoP}{\dot{\varrho}_{\scriptscriptstyle\cl{P}}}
\newcommand{\drhodP}{\mathring{\varrho}_{\scriptscriptstyle\partial\cl{P}}}
\newcommand{\nuP}{\nu_{\scriptscriptstyle\cl{P}}}
\newcommand{\nudP}{\nu_{\scriptscriptstyle\partial\cl{P}}}
\newcommand{\dnuP}{\dot{\nu}_{\scriptscriptstyle\cl{P}}}
\newcommand{\dnudP}{\mathring{\nu}_{\scriptscriptstyle\partial\cl{P}}}
\newcommand{\chiP}{\chi_{\scriptscriptstyle\cl{P}}}
\newcommand{\chidP}{\chi_{\scriptscriptstyle\partial\cl{P}}}
\newcommand{\bschiP}{\bs{\chi}_{\scriptscriptstyle\cl{P}}}
\newcommand{\bschidP}{\bs{\chi}_{\scriptscriptstyle\partial\cl{P}}}

\newcommand{\psiP}{\psi_{\scriptscriptstyle\cl{P}}}
\newcommand{\psidP}{\psi_{\scriptscriptstyle\partial\cl{P}}}

\newcommand{\dpsiP}{\dot{\psi}_{\scriptscriptstyle\cl{P}}}
\newcommand{\dpsidP}{\mathring{\psi}_{\scriptscriptstyle\partial\cl{P}}}

\newcommand{\bsbni}{\bs{b}^{\mathrm{ni}}}
\newcommand{\bsbin}{\bs{b}^{\mathrm{in}}}
\newcommand{\bsgni}{\bs{g}^{\mathrm{ni}}}
\newcommand{\bsgin}{\bs{g}^{\mathrm{in}}}

\newcommand{\bsvP}{\bs{\upsilon}_{\scriptscriptstyle\cl{P}}}
\newcommand{\bsvdP}{\bs{\upsilon}_{\scriptscriptstyle\partial\cl{P}}}
\newcommand{\bsvddP}{\bs{\upsilon}_{\scriptscriptstyle\partial^2\cl{P}}}
\newcommand{\dbsvP}{\dot{\bs{\upsilon}}_{\scriptscriptstyle\cl{P}}}
\newcommand{\dbsvdP}{\mathring{\bs{\upsilon}}_{\scriptscriptstyle\partial\cl{P}}}
\newcommand{\bsyP}{\bs{y}_{\scriptscriptstyle\cl{P}}}
\newcommand{\bsydP}{\bs{y}_{\scriptscriptstyle\partial\cl{P}}}
\newcommand{\bsxP}{\bs{x}_{\scriptscriptstyle\cl{P}}}
\newcommand{\bsxdP}{\bs{x}_{\scriptscriptstyle\partial\cl{P}}}

\newcommand{\vphiP}{\varphi_{\scriptscriptstyle\cl{P}}}
\newcommand{\vphidP}{\varphi_{\scriptscriptstyle\partial\cl{P}}}
\newcommand{\vphiddP}{\varphi_{\scriptscriptstyle\partial^2\cl{P}}}
\newcommand{\dvphiP}{\dot{\varphi}_{\scriptscriptstyle\cl{P}}}

\newcommand{\dvphidP}{\mathring{\varphi}_{\scriptscriptstyle\partial\cl{P}}}

\newcommand{\dvphiddP}{\dot{\varphi}_{\scriptscriptstyle\partial^2\cl{P}}}

\newcommand{\bsMP}{\bs{M}_{\mskip-4mu\scriptscriptstyle\cl{P}}}

\newcommand{\bsMdP}{\bs{M}_{\mskip-4mu\scriptscriptstyle\partial\cl{P}}}

\newcommand{\mP}{m_{\mskip-2mu\scriptscriptstyle\cl{P}}}
\newcommand{\mdP}{m_{\mskip-2mu\scriptscriptstyle\partial\cl{P}}}

\newcommand{\sP}{s_{\scriptscriptstyle\cl{P}}}
\newcommand{\sdP}{s_{\scriptscriptstyle\partial\cl{P}}}

\newcommand{\muP}{\mu_{\scriptscriptstyle\cl{P}}}
\newcommand{\mudP}{\mu_{\scriptscriptstyle\partial\cl{P}}}
\newcommand{\muddP}{\mu_{\scriptscriptstyle\partial^2\cl{P}}}

\newcommand{\bsjP}{\bs{\jmath}_{\scriptscriptstyle\cl{P}}}

\newcommand{\bsjdP}{\bs{\jmath}_{\scriptscriptstyle\partial\cl{P}}}

\newcommand{\prhodP}{\partial_{\rhodP}}
\newcommand{\pelldP}{\partial_{\ell_\tau}}
\newcommand{\pvphiP}{\partial_{\vphiP}}
\newcommand{\pgvphiP}{\partial_{\Grad\vphiP}}
\newcommand{\pvphidP}{\partial_{\vphidP}}
\newcommand{\pgvphidP}{\partial_{\Grads\vphidP}}

\newcommand{\DP}{\br{D}_{\mskip-2mu\scriptscriptstyle\prt}}
\newcommand{\DkP}{\br{D}^{\mathrm{0}}_{\mskip-2mu\scriptscriptstyle\prt}}
\newcommand{\DdP}{\br{D}_{\mskip-2mu\scriptscriptstyle\dprt}}
\newcommand{\DkdP}{\br{D}^{\mathrm{0}}_{\mskip-2mu\scriptscriptstyle\dprt}}

\newcommand{\brT}{\br{T}}

\newcommand{\brTvis}{{\br{T}}^{\mathrm{vis}}_{}}

\newcommand{\brH}{\br{H}}

\newcommand{\brHvis}{{\br{H}}^{\mathrm{vis}}_{}}

\newcommand{\bsxi}{\bs{\xi}}

\begin{document}

\title[A bulk-surface continuum theory for fluid flows and phase segregation]{A bulk-surface continuum theory for fluid flows and phase segregation with finite surface thickness}
\author{Anne Boschman, Luis Espath, \& Kristoffer G. van der Zee}
\address{School of Mathematical Sciences, University of Nottingham, Nottingham, NG7 2RD, United Kingdom}
\email{luis.espath@nottingham.ac.uk}

\date{\today}

\begin{abstract}
\noindent
In this continuum theory, we propose a mathematical framework to study the mechanical interplay of bulk-surface materials undergoing deformation and phase segregation. To this end, we devise a principle of virtual powers with a bulk-surface dynamics, which is postulated on a material body $\prt$ where the boundary $\dprt$ may lose smoothness, that is, the normal field may be discontinuous on an edge $\ddprt$. The final set of equations somewhat resemble the Navier--Stokes--Cahn--Hilliard equation for the bulk and the surface. Aside from the systematical treatment based on a specialized version of the virtual power principle and free-energy imbalances for bulk-surface theories, we consider two additional ingredients: an explicit dependency of the apparent surface density on the surface thickness and mixed boundary conditions for the velocity, chemical potential, and microstructure.
\\
\textbf{AMS subject classifications:}
$\cdot$
74N20 
$\cdot$
80A22 
$\cdot$
80A17 
$\cdot$
82C26 
$\cdot$
35L65 
$\cdot$

\end{abstract}

\maketitle

\tableofcontents                        


\section{Introduction}

Bulk-surface models have been used to describe a wide range of phenomena, from emulsions, foams stabilized by surface-active agents, to biological cell dynamics governed by proteins. Key to these models is the idea that the dynamics are not solely restricted to a bulk material, but that an active surface coating the bulk material, that is, a surface with its own dynamics, also dictates the overall material behaviour. Typically, these systems involve the adsorption of species onto the surface, giving rise to the particular dynamics found on the surface. These interactions may result in motion of the surface, which in turn may provoke bulk material deformation and the other way around. This bulk-surface reciprocal interplay is the focus of this work.
 
From a historical standpoint, a sharp interface of two-phase flow relates the surface traction at the interface with surface tension and curvature via a jump condition. Boussinesq \cite{Bou13} stipulated that a surface viscosity has to be incorporated in the interfacial constitutive law. As acknowledged by Bothe and Pr{\"u}ss \cite{Bot10}, Levich \cite{Lev62} also claimed that interfacial stresses may be induced by surface tension gradients due to the presence of surface-active particles, known as surfactants. The combination of such phenomena along with the mechanics of these material surfaces have been studied extensively. Adam \cite{Ada41}, Adamson \cite{Ada67}, and Scriven \cite{Scr60} were the pioneers in the physics and thermochemistry of material surfaces, as acknowledged by Gurtin \& Murdoch \cite{Gur75}; who also recast and systematically derived the underlying rational mechanics of bulk-surfaces materials.

Motivation for these bulk-surface continuum theories can for instance be found in biological systems. Cells may mathematically be described as a bulk material (cytoplasm) enclosed by a surface (cell membrane). Their mechanobiology involves various complex processes, as Ladoux \& M{\`e}ge \cite[Figure 1a]{Lad17} illustrate, which in turn determine the shape of these cells. In particular, during adhesion cells may take saucer-shaped forms. To fully understand the underlying mechanical behavior of these cells and such adhesive processes, the tractions developed on the edges need to be accounted for. Considering arbitrary geometries, including those with a boundary that may lose its smoothness, requires certain modifications and further generalization of continuum theories, as shown by Espath \cite{Esp21c} for the Navier-Stokes equations. Furthermore, Brangwynne \cite{Bra15} and Shin \& Brangwynne \cite{Shi17} suggest that membrane-less organelles are formed by regulated phase-segregation processes within the cytoplasm. In these works, the authors capitalize on the physics of polymer phase separation. Current frameworks that may capture the dynamics of these biological cells to some extent include the work by Madzvamuse et al. \cite{Mad15}, who present a reaction-diffusion for bulk-surface systems suitable to model cell polarization but do not include phase segregation and motion. In a similar fashion, Duda et al. \cite{Dud23} investigate bulk-surface systems for cell adsorption/desorption and chemical reactions for classical diffusion.

The objective of this work is to devise a continuum theory for bulk-surface materials undergoing deformation and phase segregation. In particular, we consider an immiscible binary bulk fluid enclosed by a thin immiscible binary fluid film that both deform in an incompressible manner. We treat this thin film as a material surface with finite thickness. Moreover, we assume that the material surface may lose smoothness, which gives rise to additional geometric contributions in the mathematical formulation. We depart by considering the motion of the bulk-surface material. Both bulk fluid and enclosing film of fluid undergo isochoric motions, that is, both flows are incompressible. Isochoric motion within the bulk implies no change in volume. However, isochoric motion within the thin film does not imply no change in surface area since the thickness of the surface may change. Based on these hypotheses, we derive the mass balance equations for the bulk and surface. This bulk-surface mass balance formalism is discussed in Section \S\ref{sc:mass.balance}. Next, we extend the work by Espath \cite{Esp23} and present the coupled bulk-surface principle of virtual powers in Section \S\ref{sc:pvp}. In Section \S\ref{sc:balances}, we present the partwise balance laws of forces, microforces, torques, and microtorques for the bulk-surface system, and in Section \S\ref{sc:conserved.species.transport} we account for the transport of species. Then, in Sections \S\ref{sc:free.energy.imbalance} and \S\ref{sc:constitutive.relations}, we present the thermodynamics of the system and its implications in terms of constitutive equations. Section \S\ref{sc:specialization} presents the specialized equations for a bulk-surface system undergoing phase segregation based on a particular choice for the free-energy densities. Lastly, in Section \S\ref{sc:boundary}, we present a set of boundary conditions that allow slip between the surface and the bulk with a dissipative nature, and employ these to derive the Lyapunov decay relation, which characterizes the dissipative nature of our bulk-surface system and its interaction with the environment.

\subsection{Differential tools}

Consider a smooth surface $\srf$ oriented by the outward unit normal $\bs{n}$ at $\bs{x}\in\srf$. Next, consider the smooth scalar, vector, and tensor fields defined on $\srf$, which we denote by $\kappa$, $\bs{\kappa}$, and $\br{K}$, respectively. In what follows, we define the differential operators. Bear in mind that we allow the fields $\kappa$, $\bs{\kappa}$, and $\br{K}$ to have smooth normal extensions, enabling us to define the relevant differential operators in a neighborhood of $\srf$ along all directions.

Bulk gradients may be written in the form
\begin{equation}\label{eq:gradient.scalar.surface.extension}
\Grad\kappa=\partial_n\kappa\mskip3mu\bs{n}+\partial_p\kappa\mskip3mu\bs{e}^p, \qquad \text{and} \qquad \Grad\bs{\kappa}=\partial_n\bs{\kappa}\otimes\bs{n}+\partial_p\bs{\kappa}\otimes\bs{e}^p, \qquad \text{with} \qquad p=1,2,
\end{equation}
where the contravariant bases $\bs{e}^p$ are tangential to $\srf$ and defined by the covariante bases $\bs{e}_p=\partial_p\bs{x}$ for all $\bs{x}\in\srf$ along with $\bs{e}_p \cdot \bs{e}^q = \delta_p^q$, where $\delta_p^q$ is the Kronecker delta. Next, let $\Prj{\bs{n}}\coloneqq\Prj{\bs{n}}(\bs{n})$ denote the projector onto the plane defined by $\bs{n}$ at $\bs{x}\in\srf$, which reads
\begin{equation}\label{eq:tan.projector}
\Prj{\bs{n}}\coloneqq\id-\bs{n}\otimes\bs{n}=\Prj{\bs{n}}^{\trans},
\end{equation}
where $(\cdot)^{\trans}$ represents the transposition. Then, in view of the expression \eqref{eq:gradient.scalar.surface.extension} along with \eqref{eq:tan.projector}, surface gradients are given by
\begin{equation}\label{eq:surface.gradient}
\Grads\kappa \coloneqq \partial_p\kappa\mskip3mu\bs{e}^p=\Prj{\bs{n}} \Grad\kappa, \qquad \text{and} \qquad \Grads\bs{\kappa} \coloneqq \partial_p\bs{\kappa}\otimes\bs{e}^p= (\Grad\bs{\kappa})\Prj{\bs{n}}, \qquad \text{with} \qquad p=1,2,
\end{equation}
and surface divergences by
\begin{equation}\label{eq:surface.divergence}
\Divs\bs{\kappa} \coloneqq \partial_p\bs{\kappa}\cdot\bs{e}^p=\Grad\bs{\kappa}\twovdots\Prj{\bs{n}}, \qquad \text{and} \qquad \Divs\br{K} \coloneqq \partial_p\br{K}\cdot\bs{e}^p=\Grad\br{K}\twovdots\Prj{\bs{n}} \qquad \text{with} \qquad p=1,2.
\end{equation}
Also, Laplace--Beltrami operators may be written as
\begin{equation} \label{eq:laplace.beltrami}
\triangles\kappa \coloneqq \Divs\Grads\kappa = \Grad(\Prj{\bs{n}}\Grad\kappa)\twovdots\Prj{\bs{n}}, \qquad \text{and} \qquad \triangles\bs{\kappa} \coloneqq \Divs\Grads\bs{\kappa} = \Grad((\Grad\bs{\kappa})\Prj{\bs{n}})\twovdots\Prj{\bs{n}}.
\end{equation}
Next, on a smooth closed oriented surface $\srf$ for the smooth vector and tensor fields $\bs{\kappa}$ and $\br{K}$, the surface divergence theorem states that
\begin{equation}\label{eq:smooth.divs.theo.closed.S}
\int\limits_{\srf}\Divs(\Prj{\bs{n}} \bs{\kappa})\da = 0, \qquad \text{and} \qquad \int\limits_{\srf}\Divs(\br{K}\Prj{\bs{n}})\da = \bs{0},
\end{equation}
whereas, for any smooth vector and tensor fields $\bs{\kappa}$ and $\br{K}$ on a smooth open oriented surface $\srf$, the surface divergence theorem reads
\begin{equation}\label{eq:smooth.divs.theo.open.S}
\int\limits_{\srf}\Divs(\Prj{\bs{n}} \bs{\kappa})\da=\int\limits_{\dsrf}\bs{\kappa}\cdot\bs{\nu}\ds, \qquad \text{and} \qquad \int\limits_{\srf}\Divs(\br{K}\Prj{\bs{n}})\da=\int\limits_{\dsrf}\br{K}\bs{\nu}\ds,
\end{equation}
with $\bs{\nu}$ being the outward unit tangent-normal at boundary $\dsrf$.

Lastly, consider a nonsmooth oriented surface $\srf$ for which smoothness of the normal vector field is lost on an edge $\edg$. The edge $\edg$ is defined by the limiting outward unit tangent-normals $\bs{\nu}^+$ and $\bs{\nu}^-$, yet we only consider smooth $\edg$. The surface $\srf$ has to be understood as the union of open sets $\srf \coloneqq \srf^+ \cup \srf^-$ or in general, $\srf \coloneqq \bigcup_{\alpha} \srf_\alpha$. Additionally, we abandon the smooth hypotheses of $\bs{\kappa}$ and $\br{K}$ and allow these fields to be discontinuous across $\edg$, and denote by $\bs{\kappa}^\pm$ and $\br{K}^\pm$, respectively, the limiting values of $\bs{\kappa}$ and $\br{K}$ when approaching $\edg$ from $\srf^{\pm}$. Owing to the lack of smoothness on the edge $\edg$, the surface divergence theorem exhibits a \emph{surplus}, that is,
\begin{equation}\label{eq:nonsmooth.divs.theo.closed.S}
\int\limits_{\srf}\Divs(\Prj{\bs{n}} \bs{\kappa})\da=\int\limits_{\edg}\surp{\bs{\kappa}\cdot\bs{\nu}}\ds,\qquad\text{and}\qquad\int\limits_{\srf}\Divs(\br{K}\Prj{\bs{n}})\da=\int\limits_{\edg}\surp{\br{K}\bs{\nu}}\ds,
\end{equation}
where $\surp{\bs{\kappa}\cdot\bs{\nu}}\coloneqq\bs{\kappa}^+\cdot\bs{\nu}^++\bs{\kappa}^-\cdot\bs{\nu}^-$ and $\surp{\br{K}\bs{\nu}}\coloneqq\br{K}^+\bs{\nu}^++\br{K}^-\bs{\nu}^-$. Conversely, for open nonsmooth surfaces, the surface divergence theorem \eqref{eq:nonsmooth.divs.theo.closed.S} reads
\begin{equation}\label{eq:nonsmooth.divs.theo.open.S}
\int\limits_{\srf}\Divs(\Prj{\bs{n}} \bs{\kappa})\da=\int\limits_{\dsrf}\bs{\kappa}\cdot\bs{\nu}\ds+\int\limits_{\edg}\surp{\bs{\kappa}\cdot\bs{\nu}}\ds,\qquad\text{and}\qquad\int\limits_{\srf}\Divs(\br{K}\Prj{\bs{n}})\da=\int\limits_{\dsrf}\br{K}\bs{\nu}\ds+\int\limits_{\edg}\surp{\br{K}\bs{\nu}}\ds.
\end{equation}

\section{Isochoric motion \& mass balance}
\label{sc:mass.balance}

\begin{figure}[h]
\centering
\includegraphics[width=0.65\textwidth]{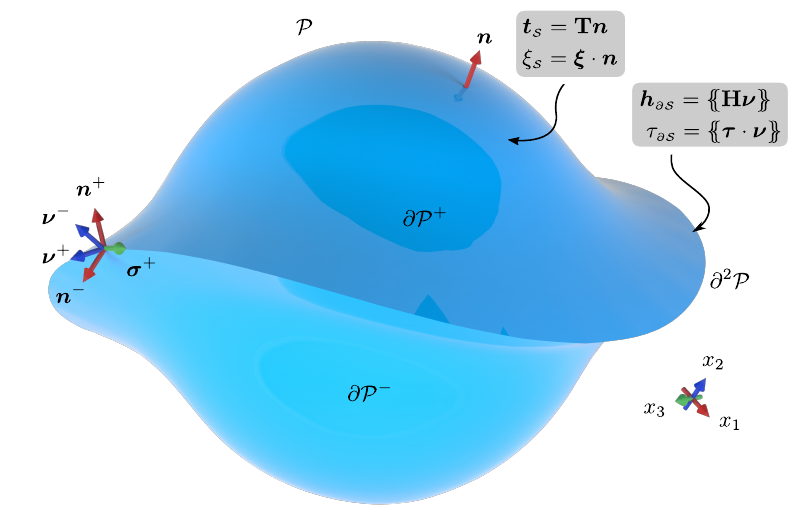}
\caption{(Adapted from \cite{Esp23}, licensed under CC-BY 4.0). Part $\prt$ with nonsmooth boundary surface $\dprt^\pm$ oriented by the unit normal $\bs{n}$ with the outward unit tangent-normal $\bs{\nu}^\pm$ on the smooth boundary-edge $\ddprt$ oriented by the unit tangent $\bs{\sigma}\coloneqq\bs{n}\times\bs{\nu}$. The surface $\dprt$ lacks smoothness on an edge $\ddprt$.}
\label{fg:geometry}
\end{figure}

In the reference configuration, we consider a material body $\prt$ occupying a three-dimensional point space $\cl{E}$. Here, geometry is arbitrary in the following sense. The closed surface boundary of the part $\prt$, denoted by $\dprt$, may lose smoothness along a curve, namely an edge $\ddprt$, see Figure \ref{fg:geometry}. In a neighborhood of an edge $\ddprt$, two smooth surfaces $\dprt^{\pm}$ are defined. The limiting unit normals of $\dprt^{\pm}$ on $\ddprt$ are denoted by the pair $\{\bs{n}^+,\bs{n}^-\}$. The pair of unit normals characterizes the edge $\ddprt$. Similarly, the limiting outward unit tangent-normal of $\dprt^{\pm}$ on $\ddprt$ are $\{\bs{\nu}^+,\bs{\nu}^-\}$. Additionally, $\ddprt$ is oriented by the unit tangent $\bs{\sigma}\coloneqq\bs{\sigma}^+$ such that $\bs{\sigma}^+\coloneqq\bs{n}^+\times\bs{\nu}^+$. As notational agreement, we use the subscript $\tau$ to refer to spatial entities in the current configuration. More specifically, we let $\prt_\tau$ denote a spatial part, we write $\dprt_\tau$ for the boundary of $\prt_\tau$ and use $\ddprt_\tau$  for the boundary of $\dprt_\tau$. Note that $\prt$, $\dprt$ and $\ddprt$ are reserved for their counterparts in the reference configuration, respectively.

In this continuum theory, the bulk and surface material are endowed with two distinct kinematic descriptors, namely, the fluid velocities $\bsvP$ in the bulk and $\bsvdP$ on the surface. Furthermore, we let $\rhoP$ be the density of the bulk material and $\rhodP$ the apparent surface density of the surface material endowed with a finite thickness $\ell_\tau$. 

In what follows, we assume that
\begin{enumerate}[label=(A.\arabic*)]
\item
The `bulk-surface motion' is isochoric. For the bulk material, this means that motion preserves its volumes. For the surface material, we assume that isochoric motion implies volume conservation on a microscopic scale. \\[4pt] \label{as:1}
\item The normal components of the velocities are continuous\footnote{ Note that we do not require continuity of the tangential velocities $\Prj{\bs{n}}\bsvP\big|_{\dprt}$ and $\Prj{\bs{n}}\bsvdP$.}, that is,
\begin{equation}\label{eq:continuous.normal.v}
\bsvP\cdot\bs{n}\big|_{\dprt} = \bsvdP\cdot\bs{n}.
\end{equation}\label{as:2}
\end{enumerate}

The kinematic constraint \eqref{eq:continuous.normal.v} is imposed to guarantee that the surface and the bulk material's boundary coincide at all times and never detach from one another. Conversely, we do not endow edges with their own kinematic descriptors. Thus, it is not required to define additional kinematic constraints. Moreover, as a consequence of the kinematic constraint \eqref{eq:continuous.normal.v} on $\ddprt$, we find that
\begin{equation}\label{eq:continuous.normal.v.consequence}
\bsvP\cdot\bs{n}^{+}\big|_{\ddprt} = \bsvdP\cdot\bs{n}^{+}\big|_{\ddprt}, \quad \text{and} \quad \bsvP\cdot\bs{n}^-\big|_{\ddprt} = \bsvdP\cdot\bs{n}^-\big|_{\ddprt}.
\end{equation}

With these assumptions in mind, let
\begin{equation}
\brFP \coloneqq \Grad^{\!\bs{x}} \bsyP, \quad \text{and} \quad \brFdP \coloneqq \Grads^{\!\bs{x}} \bsydP,
\end{equation}
denote the bulk and surface deformation gradient with respect to the reference configuration, either $\bsxP \in \prt$ or $\bsxdP \in \dprt$, respectively. Here, $\bsyP$ represents the motion of $\prt_{\tau}$ and $\bsydP$ the motion of $\dprt_{\tau}$ such that $\prt_\tau = \bsyP (\prt)$ and $\dprt_\tau = \bsydP (\dprt)$, respectively. If not made explicit, differential operators are computed with respect to the current configuration, either $\bsyP$ or $\bsydP$. Additionally, the dot operator denotes the bulk material time-derivative, that is, $\dot{s} \coloneqq \partial_t s + \bsvP \cdot \Grad s$, where $\partial_t$ is the conventional time partial derivative. Conversely, the ring operator represents the surface material derivative, derived by Cermelli et al. \cite{Cer05}, and is given by 
\begin{equation}
\mathring{s} \coloneqq \sqr{s} + \bsvdP \cdot \Grads s,
\end{equation}
where the normal time derivative is
\begin{equation}
\sqr{s} \coloneqq \dfrac{\text{d}}{\text{d}\varepsilon}\Big(s(\bsydP+\varepsilon(\bsvdP(\bsydP,t)\cdot\bs{n}(\bsydP))\bs{n}(\bsydP),t+\varepsilon)\Big)\Big|_{\varepsilon=0}.
\end{equation}

Since $\brFdP$ is rank deficient, we introduce the pseudo-inverse of the surface deformation gradient as
\begin{equation}
\brFdP^{-1} \coloneqq \Prj{n}(\bsxdP)(\brFdP + \bs{n}(\bsydP) \otimes \bs{n}(\bsxdP))^{-1}, 
\end{equation}
where we explicitly show the dependency on either the reference configuration $\bsxdP \in \dprt$ or the current configuration $\bsydP \in \dprt_{\tau}$. Note that $\brFdP + \bs{n}(\bsydP) \otimes \bs{n}(\bsxdP)$ is full rank since $\bs{n}(\bsydP)$ is not in the range of $\brFdP$. Additionally, we have that 
\begin{equation}
\brFdP^{-1} \brFdP = \Prj{n}(\bsxdP), \quad \brFdP \brFdP^{-1} = \Prj{n}(\bsydP), \quad \text{and} \quad \brFdP^{-1} \bs{n}(\bsydP) = \brFdP^{-\trans} \bs{n}(\bsxdP) = \bs{0}.
\end{equation}
In the rest of this work, all quantities depend on $\bsydP$; therefore, we drop all the arguments. 
For further details on this pseudo-inverse, the interested reader is referred to the work by \v{S}ilhav\'y \cite{Sil13} on interactions of shells with bulk matter, and also to the work by Tomassetti \cite{Tom23} on a coordinate-free description for thin
shells.

Next, we introduce $\brLP \coloneqq \Grad \bsvP$ and $\brLdP \coloneqq \Grads \bsvdP$ as the bulk and surface velocity gradient, respectively. We consider the following classical identity in continuum mechanics by Gurtin \cite{Gur82}, also known as Jacobi's formula, 
\begin{equation}\label{eq:id.bulk.1}
\dot{\overline{|\brFP|}} = |\brFP| \tr(\dbrFP \brFP^{-1}),
\end{equation}
where $|\brFP|$ denotes the determinant of $\brFP$.
The surface counterpart of expression \eqref{eq:id.bulk.1} reads
\begin{equation}\label{eq:id.surf.1}
\mathring{\overline{|\brFdP|}} = |\brFdP| \tr(\dbrFdP \brFdP^{-1}).
\end{equation}
Furthermore, we have the volumetric and areal Jacobian of deformation, respectively, defined as
\begin{equation}\label{eq:jacobian.vol}
\JP \coloneqq \dfrac{\dv_\tau}{\dv} = |\brFP|, 
\end{equation}
and
\begin{equation}\label{eq:jacobian.area}
\JdP \coloneqq \dfrac{\da_\tau}{\da} = |\brFdP|,
\end{equation}
where $\dv$ and $\dv_\tau$ are the differential of volume at the reference and current configurations, respectively. Finally, in view of \eqref{eq:id.bulk.1} and \eqref{eq:id.surf.1} and bearing in mind that $\dbrFP = \brLP \brFP$ and $\dbrFdP = \brLdP \brFdP$, we obtain
\begin{equation}\label{eq:bulk.jacobian.derivative}
\begin{aligned}
\dJP &= \JP \mskip3mu \tr(\dbrFP \brFP^{-1}), \\[4pt]
{} &= \JP \mskip3mu \tr(\brLP \brFP \brFP^{-1}), \\[4pt]
{} &= \JP \mskip3mu \Div \bsvP,
\end{aligned}
\end{equation}
where $\text{div}$ is the bulk divergence, and
\begin{equation}\label{eq:surf.jacobian.derivative}
\begin{aligned}
\dJdP &= \JdP \mskip3mu \tr(\dbrFdP \brFdP^{-1}), \\[4pt]
{} &= \JdP \mskip3mu \tr(\brLdP \brFdP \brFdP^{-1}), \\[4pt]
{} &= \JdP \mskip3mu \Divs \bsvdP.
\end{aligned}
\end{equation}

Next, following Assumption \ref{as:1}, we consider the isochoric motion
\begin{equation}\label{eq:isochoric}
\dot{\overline{|\prt_\tau|}} \coloneqq \dot{\overline{\mathrm{vol}(\prt_\tau)}}=0, 
\end{equation}
and assuming that expression \eqref{eq:isochoric} holds for any $\cl{R}_\tau \subseteq \prt_\tau$ with expression \eqref{eq:bulk.jacobian.derivative}, we are led to
\begin{align}\label{eq:volume.integral}
0 = \dot{\overline{\intRt\dv_\tau}} &= \intR\dJP\dv, \nonumber\\[4pt]
&= \intRt\Div\bsvP\dv_\tau.
\end{align}
Thus, it follows by localization
\begin{equation}\label{eq:isochoric.constraint.bulk}
\Div\bsvP = 0, \qquad \text{in } \prt_\tau.
\end{equation}

Now, using the transport theorem, the partwise bulk balance of mass is given by
\begin{equation}\label{eq:partiwise.bulk.mass}
\dot{\overline{\intRt\rhoP\dv_\tau}} = \intRt(\drhoP + \rhoP \mskip3mu \Div \bsvP)\dv_\tau = 0,
\end{equation}
and by localization we arrive at the pointwise bulk balance of mass
\begin{equation}\label{eq:pointwise.bulk.mass}
\drhoP + \rhoP \mskip3mu \Div \bsvP = 0, \qquad \text{in } \prt_\tau,
\end{equation}
and in terms of specific volume $\nuP \coloneqq \rhoP^{-1}$,
\begin{equation}\label{eq:pointwise.bulk.specific.volume}
\dnuP = \nuP \mskip3mu \Div \bsvP, \qquad \text{in } \prt_\tau.
\end{equation}
Using the isochoric constraint \eqref{eq:isochoric.constraint.bulk}, $\Div \bsvP = 0$, we obtain
\begin{equation}\label{eq:}
\drhoP = 0, \qquad \text{and} \qquad \dnuP=0,
\end{equation}
implying that the bulk mass density $\rhoP $ and specific bulk volume $\nuP$ are independent of time. Moreover, we further restrict $\rhoP $ and $\nuP$ by assuming that 
\begin{equation}\label{eq:constant.bulk.density}
\rhoP = \text{constant}, \qquad \text{and} \qquad \nuP=\text{constant},
\end{equation}
which means that applications where spatial variations of the density are important, such as oceanic and mantle convection, are excluded \cite{Gur10}. 
Based on these results for the bulk, it would be natural to assume that the surface density $\rhodP$ is constant as well for all motions. However, this would imply that both the volume and surface area do not change. This hypothesis for the material motion is far too restrictive, and provides the main motivation for taking on a microscopic viewpoint for the surface material's motion.

Thus, from a microscopic standpoint, we consider that the surface fluid has a constant density $\rho$, defined as
\begin{equation}\label{eq:constant.rho}
\rho \coloneqq \dfrac{\text{d}m_\tau}{\dv_\tau} = \text{constant},
\end{equation}
where $\text{d}m_\tau$ is the differential of mass. When the surface fluid deforms the actual thickness should change to maintain the volume constant. The apparent surface density $\rhodP$, a macroscopic quantity, is defined as
\begin{equation}\label{eq:surf.micro.macro.mass}
\rhodP \coloneqq \dfrac{\text{d}m_\tau}{\da_\tau} = \dfrac{\text{d}m_\tau}{\dv_\tau} \ell_\tau = \rho \mskip3mu \ell_\tau,
\end{equation}
where $\ell_\tau$ and $\da_\tau$ are the current thickness and current differential of area, respectively. Furthermore, let $\ell$ and $\da$ be the initial thickness and differential area, respectively. Since the microscopic surface motion is isochoric, we have that
\begin{equation}\label{eq:isochoric.differential}
\da_\tau\times\ell_\tau=\da\times\ell,
\end{equation}
leading us to a microscopic isochoric motion
\begin{equation}\label{eq:micro.isochoric}
\dot{\overline{|\ell_\tau\dprt_\tau|}} \coloneqq \dot{\overline{\mathrm{area}(\ell_\tau\dprt_\tau)}}=0.
\end{equation}
Moreover, assuming that expression \eqref{eq:micro.isochoric} holds for any $\srf_\tau \subseteq \dprt_\tau$, with expression \eqref{eq:surf.jacobian.derivative}, we have that the partwise surface balance of mass reads
\begin{align}
0 &= \dot{\overline{\int\limits_{\srf_\tau} \ell_\tau \da_\tau}}, \nonumber\\[4pt]
&= \int\limits_{\srf} \mathring{\overline{\ell_\tau \JdP}} \da, \nonumber\\[4pt]
&= \int\limits_{\srf} (\mathring{\ell}_\tau \JdP + \ell_\tau \dJdP) \da, \nonumber\\[4pt]
&= \int\limits_{\srf_\tau} (\mathring{\ell}_{\tau} + \ell_{\tau} \Divs \bsvdP) \da_\tau,
\end{align}
which, by localization, renders the pointwise surface balance of mass
\begin{equation}\label{eq:ell.balance}
\mathring{\ell}_{\tau} + \ell_{\tau} \mskip3mu \Divs \bsvdP = 0, \qquad \text{on } \dprt_\tau.
\end{equation}

In view of \eqref{eq:surf.micro.macro.mass}, we have that
\begin{equation}\label{eq:density.relation}
\dfrac{\rhodP}{\ell_\tau} = \rho, \qquad \rhodP \times \da_\tau = \rho \times \ell \times \da, \qquad\text{and}\qquad \nu \times \da_\tau = \nudP \times \ell \times \da,
\end{equation}
where $\nu \coloneqq \rho^{-1}$ is the constant microscopic specific volume and $\nudP \coloneqq \rhodP^{-1} $ is the apparent specific area. Moreover, \eqref{eq:density.relation} yields
\begin{equation}\label{eq:surface.density.constraint}
\nudP = \dfrac{\JdP}{\ell} \nu.
\end{equation}
Additionally, in view of \eqref{eq:isochoric.differential} and the areal Jacobian \eqref{eq:jacobian.area}, we have
\begin{equation}\label{eq:surf.jacobian.thickness}
\JdP=\dfrac{\ell}{\ell_\tau}.
\end{equation}
Next, multiplying \eqref{eq:ell.balance} by $\rho$ and using \eqref{eq:density.relation}$_{1}$, we arrive at
\begin{equation}\label{eq:surf.mass.balance}
\drhodP + \rhodP \mskip3mu \Divs \bsvdP = 0, \qquad \text{on } \dprt_\tau,
\end{equation}
where $\Divs \bsvdP = \Divs (\Prj{\bs{n}} \bsvdP) - 2 K \mskip3mu \bsvdP \cdot \bs{n}$ with $K \coloneqq - \fr{1}{2} \Divs \bs{n}$ being the mean curvature. Notice that one can also arrive at the pointwise surface balance of mass \eqref{eq:surf.mass.balance} by formulating a partwise balance of mass for spatially convecting regions $\cl{S}_\tau \subseteq \dprt_\tau$ and using the localization argument. We wish to emphasize that the definition for $\rho$ \eqref{eq:constant.rho} is consistent with \eqref{eq:surf.mass.balance}, as it can be written as $\mathring{\rho} \ell_\tau=0$, and we restrict our attention to applications where the macroscopic surface density is independent of spatial variations. Also, expression \eqref{eq:surf.mass.balance}, may be written in terms of the apparent specific area, that is,
\begin{equation}\label{eq:surf.specific.volume.balance}
\dnudP = \nudP (\Divs (\Prj{\bs{n}} \bsvdP) - 2 K \mskip3mu \bsvdP \cdot \bs{n}), \qquad \text{on } \dprt_\tau.
\end{equation}

Lastly, for bulk and surface fields $\phiP$ and $\phidP$ the combination of the Reynolds' transport theorems with the pointwise balances \eqref{eq:pointwise.bulk.mass} and \eqref{eq:ell.balance} renders the following identities
\begin{align}\label{eq:id.reynolds.mass.P}
\dot{\overline{\intRt \rhoP \phiP \dv_\tau}} &= \intRt (\dot{\overline{\rhoP \phiP}} + \rhoP \phiP \mskip3mu \Div \bsvP) \dv_\tau, \nonumber \\[4pt]
&= \intRt (\rhoP \dphiP + (\drhoP + \rhoP \mskip3mu \Div \bsvP) \phiP) \dv_\tau, \nonumber \\[4pt]
&= \intRt \rhoP \dphiP \dv_\tau,
\end{align}
and, bearing in mind that $\rhodP = \rho \ell_\tau$, we have that
\begin{align}\label{eq:id.reynolds.mass.dP}
\dot{\overline{\int\limits_{\srf_\tau} \rhodP \phidP \da_\tau}} &= \dot{\overline{\int\limits_{\srf_\tau} \rho \ell_\tau \phidP \da_\tau}} \nonumber \\[4pt]
&= \int\limits_{\srf_\tau} \rho  (\mathring{\overline{\ell_\tau \phidP}} + \ell_\tau \phidP \mskip3mu \Divs \bsvdP) \da_\tau, \nonumber \\[4pt]
&= \int\limits_{\srf_\tau} \rho (\ell_\tau \dphidP + (\mathring{\ell}_{\tau} + \ell_{\tau} \mskip3mu \Divs \bsvdP) \phidP) \da_\tau, \nonumber \\[4pt]
&= \int\limits_{\srf_\tau} \rho \ell_\tau \dphidP \da_\tau.
\end{align}

\section{Virtual power principle}
\label{sc:pvp}

To derive the field equations of this continuum framework, we devise the following principle of virtual powers based on the work by Espath \cite{Esp23}, where we consider the presence of scalar and vectorial virtual fields in the bulk as well as on the surface. That is, the principle of virtual powers for bulk-surface materials undergoing motion and phase segregation reads
\begin{equation}\label{eq:virtual.power.principle}
\cl{V}_{\mathrm{ext}}(\prt_\tau,\dprt_\tau;\bschiP,\chiP,\bschidP,\chidP) = \cl{V}_{\mathrm{int}}(\prt_\tau,\dprt_\tau;\bschiP,\chiP,\bschidP,\chidP),
\end{equation}
where $\chiP$ and $\bschiP$ are, respectively, sufficiently smooth scalar and vector virtual fields defined in $\prt_\tau$, and similarly, $\chidP$ and $\bschidP$ are, respectively, sufficiently smooth scalar and vector virtual fields defined on $\dprt_\tau$. The external and internal virtual power are, respectively, given by
\begin{align}\label{eq:external.virtual.power}
\cl{V}_{\mathrm{ext}}(\prt_\tau,\dprt_\tau;\bschiP,\chiP,\bschidP,\chidP) \coloneqq{}& \intPt \bs{b} \cdot \bschiP \dv_\tau + \intdPt (\bs{g} - \bsts) \cdot \bschidP \da_\tau + \intdPt \bsts \cdot \bschiP \da_\tau + \intddPt \bshds \cdot \bschidP \ds_\tau \nonumber\\[4pt]
&+ \intPt \gamma \chiP \dv_\tau + \intdPt (\zeta - \xis) \chidP \da_\tau + \intdPt \xis \chiP \da_\tau + \intddPt \tauds \chidP \ds_\tau,
\end{align}
and
\begin{align}\label{eq:internal.virtual.power}
\cl{V}_{\mathrm{int}}(\prt_\tau,\dprt_\tau;\bschiP,\chiP,\bschidP,\chidP) \coloneqq{}& \intPt \br{T} \twovdots \Grad \bschiP \dv_\tau + \intdPt \br{H} \twovdots \Grads\bschidP \da_\tau \nonumber\\[4pt]
&+ \intPt \bs{\xi} \cdot \Grad \chiP \dv_\tau - \intPt \pi \chiP \dv_\tau + \intdPt \bs{\tau} \cdot \Grads\chidP \da_\tau - \intdPt \varpi \chidP \da_\tau.
\end{align}
Here, entering in the external virtual power, $\bs{b}$ is the bulk external force, $\bs{g}$ is the surface external force, $\bsts$ is the surface traction, $\bshds$ is the edge traction, $\gamma$ is the bulk external microforce, $\zeta$ is the surface external microforce, $\xis$ is the surface microtraction, and $\tauds$ is the edge microtraction. Also, entering in the internal virtual power, $\br{T}$ is the bulk stress, $\br{H}$ is the surface stress, $\bs{\xi}$ is the bulk microstress, $\pi$ is the bulk internal microforce, $\bs{\tau}$ is the surface microstress, and $\varpi$ is the surface internal microforce.

Next, by combining \eqref{eq:external.virtual.power} and \eqref{eq:internal.virtual.power} through \eqref{eq:virtual.power.principle} and using the divergence theorem and the surface divergence theorem for nonsmooth closed surfaces \eqref{eq:nonsmooth.divs.theo.closed.S}, we are led to
\begin{multline}
\intPt \bschiP \cdot (\Div\br{T} + \bs{b}) \dv_\tau + \intdPt \bschiP \cdot (\bsts - \br{T} \cdot \bs{n}) \da_\tau \\[4pt]
+ \intPt \chiP (\Div\bs{\xi} + \pi + \gamma) \dv_\tau + \intdPt \chiP (\xis - \bs{\xi} \cdot \bs{n}) \da_\tau \\[4pt]
+ \intdPt \bschidP \cdot (\Divs(\br{H}\Prj{\bs{n}}) + \bs{g} - \bsts) \da_\tau + \intddPt \bschidP \cdot (\bshds - \surp{\br{H} \bs{\nu}}) \ds_\tau \\[4pt]
+ \intdPt \chidP (\Divs(\Prj{\bs{n}}\bs{\tau}) + \varpi + \zeta - \xis) \da_\tau + \intddPt \chidP (\tauds - \surp{\bs{\tau} \cdot \bs{\nu}}) \ds_\tau = 0.
\end{multline}
Then, by variational arguments, the surface and edge tractions are, respectively, given by
\begin{equation}\label{eq:tractions}
\bsts = \br{T} \bs{n}, \qquad \text{on } \dprt_\tau, \qquad \text{and} \qquad \bshds = \surp{\br{H} \bs{\nu}}, \qquad \text{on } \ddprt_\tau,
\end{equation}
while the surface and edge microtractions are, respectively, given by
\begin{equation}\label{eq:microtractions}
\xis = \bs{\xi} \cdot \bs{n}, \qquad \text{on } \dprt_\tau, \qquad \text{and} \qquad \tauds = \surp{\bs{\tau} \cdot \bs{\nu}}, \qquad \text{on } \ddprt_\tau,
\end{equation}
and the bulk and surface field equations for motion are, respectively, given by
\begin{equation}\label{eq:field.equations.fluid}
\Div\br{T} + \bs{b} = \bs{0}, \qquad \text{in } \prt_\tau, \qquad \text{and} \qquad \Divs(\br{H}\Prj{\bs{n}}) + \bs{g} - \bsts = \bs{0}, \qquad \text{on } \dprt_\tau,
\end{equation}
and the bulk and surface field equations for phase segregation are, respectively, given by
\begin{equation}\label{eq:field.equations.phase}
\Div\bs{\xi} + \pi + \gamma = 0, \qquad \text{in } \prt_\tau \qquad \text{and} \qquad \Divs(\Prj{\bs{n}}\bs{\tau}) + \varpi + \zeta - \xis = 0, \qquad \text{on } \dprt_\tau.
\end{equation}
Additionally, splitting $\Divs(\br{H}\Prj{\bs{n}}) = \Divs\br{H} + 2K \br{H} \bs{n}$ and $\Divs(\Prj{\bs{n}}\bs{\tau}) = \Divs\bs{\tau} + 2K \bs{\tau} \cdot \bs{n}$, the surface field equations \eqref{eq:field.equations.fluid}$_2$ and \eqref{eq:field.equations.phase}$_2$, may take the following form
\begin{equation}\label{eq:field.equations.fluid.mod}
\Divs\br{H} + 2K \br{H} \bs{n} + \bs{g} - \bsts = \bs{0}, \qquad \text{and} \qquad \Divs\bs{\tau} + 2K \bs{\tau} \cdot \bs{n} + \varpi + \zeta - \xis = 0, \qquad \text{on } \dprt_\tau.
\end{equation}

Lastly, we decompose the external bulk force and external surface force into an inertial and non-inertial part, respectively, that is, 
\begin{equation}\label{eq:generalised.body.force}
\bs{b} \coloneqq \bsbni + \bsbin, \qquad \text{and} \qquad \bs{g} \coloneqq \bsgni + \bsgin.
\end{equation}
For the inertial bulk and surface forces, we consider the relations
\begin{equation}\label{eq:inertial.body.force}
\bsbin \coloneqq -\rhoP \dbsvP, \qquad \text{and} \qquad \bsgin \coloneqq -\rhodP \dbsvdP = - \rho \ell_\tau \dbsvdP.
\end{equation}
Therefore, expressions \eqref{eq:field.equations.fluid} take the form
\begin{equation}\label{eq:field.equations.fluid.explicit}
\rhoP \dbsvP = \Div\br{T} + \bsbni, \qquad \text{in } \prt_\tau, \qquad \text{and} \qquad \rho \ell_\tau \dbsvdP = \Divs(\br{H}\Prj{\bs{n}}) + \bsgni - \bsts, \qquad \text{on } \dprt_\tau,
\end{equation}
Similar arguments may be used to decompose the bulk and surface external microforces into inertial and non-inertial parts.

\begin{rmk}[Motivation for discontinuous surface stress and surface microstress]
In higer-order continua, it is oftentimes found that edge tractions and edge microtractions are of the form
\begin{equation}
\bshds \coloneqq \surp{(\bb{G}\bs{n})\bs{\nu}} \qquad \text{and} \qquad \tauds \coloneqq \surp{(\bs{\Sigma} \bs{n}) \cdot \bs{\nu}},
\end{equation}
where $\bb{G}$ and $\bs{\Sigma}$ are, respectively, the hyperstress and the hypermicrostress see, for instance, Espath et al. \cite{Esp20,Esp21a} for phase field gradient theories and Fosdick \cite{Fos16} for gradient theories in solid and fluid mechanics. Therefore, although one assumes that $\bb{G}$ and $\bs{\Sigma}$ are continuous, it is expected that $\bb{G}\bs{n}^+ \neq \bb{G}\bs{n}^-$ and $\bs{\Sigma} \bs{n}^+ \neq \bs{\Sigma} \bs{n}^-$ on $\ddprt$. Conversely, in bulk-surface systems, edge tractions \eqref{eq:tractions}$_2$ and edge microtraction \eqref{eq:microtractions}$_2$ take the form
\begin{equation}
\bshds \coloneqq \surp{\br{H}\bs{\nu}} \qquad \text{and} \qquad \tauds \coloneqq \surp{\bs{\tau} \cdot \bs{\nu}},
\end{equation}
on $\ddprt$. Thus, one may argue that the surface stress $\br{H}$ and surface microstress $\bs{\tau}$ play a similar role to the hyperstress and hypermicrostress times the limiting normals, that is, $\bb{G}\bs{n}^\pm$ and $\bs{\Sigma} \bs{n}^\pm$. Therefore, we consider that the surface stress $\br{H}$ and surface microstress $\bs{\tau}$ are piecewise smooth but can be discontinuous on an edge $\ddprt$.
\end{rmk}

\subsection{Frame indifference principle}

We now require the internal virtual power to be indifferent to frame changes. That is,
\begin{equation}
\cl{V}_{\mathrm{int}}(\prt_\tau,\dprt_\tau;\bschiP,\chiP,\bschidP,\chidP) = \cl{V}_{\mathrm{int}}(\prt_\tau,\dprt_\tau;\bschiP + \bs{\beta}+\br{\Omega}\bs{y},\chiP,\bschidP+\bs{\beta}+\br{\Omega}\bs{y},\chidP),
\end{equation}
where $\bs{y} \in \prt_\tau \cup \dprt_\tau$, $\bs{\beta}$ is a constant (in space) velocity, and $\br{\Omega}$ is a constant (in space) rotation, which are applied to the frame of reference. Note that $\br{\Omega}$ is a skew-symmetric tensor.

Since the frame indifference only plays a part for vector quantities, we restrict attention to the fields $\bschiP$ and $\bschidP$. Thus, enforcing that the internal power must be indifferent to this change of frame yields
\begin{equation}
\intPt \br{T} \twovdots \Grad \bschiP \dv_\tau + \intdPt \br{H} \twovdots \Grads\bschidP \da_\tau = \intPt \br{T} \twovdots (\Grad \bschiP + \br{\Omega}) \dv_\tau + \intdPt \br{H} \twovdots (\Grads\bschidP + \br{\Omega}\Prj{\bs{n}}) \da_\tau,
\end{equation}
which may be localized to
\begin{align}\label{eq:pointwise.frame.indifference}
\br{T} \twovdots \br{\Omega} = \bs{0}, \quad \text{in } \prt_\tau, \qquad \text{and} \qquad \br{H} \twovdots \br{\Omega}\Prj{\bs{n}} = \bs{0}  \quad \text{on } \dprt_\tau.
\end{align}
Note that from \eqref{eq:pointwise.frame.indifference}$_2$, we have that
\begin{equation}\label{eq:cond.H}
\br{H} \twovdots \br{\Omega} = \br{H}\bs{n} \cdot \br{\Omega}\bs{n}, \qquad \text{on } \dprt_\tau.
\end{equation}
Since \eqref{eq:pointwise.frame.indifference} must hold for all skew-symmetric tensors, these expressions imply $\br{T} = \br{T}^{\trans}$ and $\br{H}\Prj{n} = \Prj{n}\br{H}^{\trans}$. The latter implies that both $\br{H}=\br{H}^{\trans}$ and $\br{H}\bs{n} = \bs{0}$.

\section{Partwise balance of forces, microforces, torques \& microtorques}
\label{sc:balances}

Integrating the field equations, balances of forces \eqref{eq:field.equations.fluid}, on each respective part, we have that
\begin{equation}
\intPt (\Div \br{T} + \bs{b}) \dv_\tau + \intdPt (\Divs (\br{H}\Prj{n}) + \bs{g} - \bsts) \da_\tau = \bs{0},
\end{equation}
and using the divergence theorem and the divergence theorem on nonsmooth closed surfaces \eqref{eq:nonsmooth.divs.theo.closed.S}, we arrive at
\begin{equation}
\intPt \bs{b} \dv_\tau + \intdPt \br{T}\bs{n} \da_\tau + \intdPt (\bs{g} - \bsts) \da_\tau + \intddPt \surp{\br{H}\bs{\nu}} \ds_\tau = \bs{0}.
\end{equation}
Lastly, using expressions \eqref{eq:tractions} for the surface and edge tractions, we find the partwise bulk-surface balance of forces
\begin{equation}\label{eq:partwise.balance.forces}
\cl{F}^\sharp(\prt_\tau,\dprt_\tau) \coloneqq \intPt \bs{b} \dv_\tau + \intdPt \bs{g} \da_\tau + \intddPt \bshds \ds_\tau = \bs{0}.
\end{equation}
Similarly, by emulating the above procedure for the remaining field equations, the balances of microforces \eqref{eq:field.equations.phase}, we arrive at the partwise bulk-surface balance of microforces
\begin{equation}\label{eq:partwise.balance.microforces}
\cl{F}^\flat(\prt_\tau,\dprt_\tau) \coloneqq \intPt (\pi + \gamma) \dv_\tau + \intdPt (\varpi + \zeta) \da_\tau + \intddPt \tauds  \ds_\tau = 0.
\end{equation}
Note that the surface traction and the surface microtraction do not appear in the partwise bulk-surface balance of forces \eqref{eq:partwise.balance.forces} and microforces \eqref{eq:partwise.balance.microforces} as opposed to conventional continuum mechanical theories.

To arrive at the partwise bulk-surface balances of torques and microtorques of the bulk-surface material, we need to introduce the position vector $\bs{r} \coloneqq \bs{y} - \bs{o}$, where $\bs{y} \in \prt_\tau \cup \dprt_\tau$ and an origin arbitrarily chosen and fixed $\bs{o} \in \cl{E}$. First, we derive the partwise bulk-surface balance of torques, for which we take the tensor product between $\bs{r}$ and the force balances in \eqref{eq:field.equations.fluid}. The resulting expression is integrated over the respective parts, yielding the balance
\begin{equation}\label{eq:balance.torques.1}
\intPt \bs{r} \otimes (\Div \br{T} + \bs{b}) \dv_\tau + \intdPt \bs{r} \otimes (\Divs (\br{H}\Prj{n}) + \bs{g} - \bsts) \da_\tau = \bs{0}.
\end{equation}
Noting that $\Grad \bs{r} = \id$ and $\Grads\bs{r} = \Prj{\bs{n}}$, we employ the following identities
\begin{equation}\label{eq:identities.torques.balance}
\Div(\bs{r} \otimes \br{T}) = \br{T} + \bs{r} \otimes \Div\br{T}, \qquad \text{and} \qquad \Divs(\bs{r} \otimes \br{H}\Prj{\bs{n}}) = \Prj{\bs{n}}\br{H}\Prj{\bs{n}} + \bs{r} \otimes \Divs(\br{H}\Prj{\bs{n}}),
\end{equation}
along with the divergence theorem and the divergence theorem on nonsmooth closed surfaces \eqref{eq:nonsmooth.divs.theo.closed.S}, to obtain 
\begin{equation}\label{eq:balance.torques.2}
\intPt (\bs{r} \otimes \bs{b} - \br{T} )\dv_\tau + \intdPt \bs{r} \otimes \br{T} \bs{n} \da_\tau + \intdPt (\bs{r} \otimes (\bs{g} - \bsts) - \Prj{n}\br{H}\Prj{n} ) \da_\tau + \intddPt \surp{\bs{r} \otimes \br{H}\bs{\nu}}\ds_\tau = \bs{0}.
\end{equation}
Then, using the definitions for the surface and edge tractions \eqref{eq:tractions}, we arrive at 
\begin{equation}\label{eq:balance.torques.3}
\intPt (\bs{r} \otimes \bs{b} - \br{T} )\dv_\tau + \intdPt (\bs{r} \otimes \bs{g} - \Prj{n}\br{H}\Prj{n} ) \da_\tau  + \intddPt \bs{r} \otimes \bshds \ds_\tau = \bs{0}.
\end{equation}

Lastly, by summing \eqref{eq:balance.torques.3} with its negative transposed, we obtain the partwise bulk-surface balance of torques
\begin{equation}\label{eq:balance.torques.4}
\cl{T}^\sharp(\prt_\tau, \dprt_\tau) \coloneqq \intPt \bs{r} \wedge \bs{b}\dv_\tau + \intdPt \bs{r} \wedge \bs{g} \da_\tau  + \intddPt \bs{r} \wedge \bshds \ds_\tau = \bs{0},
\end{equation}
where we used the implications of frame-indifference, that is, $\br{T} = \br{T}^{\trans}$ and $\br{H}\Prj{n} = \Prj{n}\br{H}^{\trans}$, (and consequently $\Prj{n}\br{H}\Prj{n} = \Prj{n}\br{H}^{\trans}\Prj{n}$ since $\Prj{n}^2=\Prj{n}$). Additionally, we used the wedge product defined as $\bs{a} \wedge \bs{b} \coloneqq \bs{a} \otimes \bs{b} - \bs{b} \otimes \bs{a}$.

We now construct the partwise bulk-surface balance of microtorques. As opposed to the balance of torques, balance of microtorques cannot be presented as the sum of wedge products. First, we multiply the microforces balances in \eqref{eq:field.equations.phase} by the position vector $\bs{r}$, and then integrate the results over the respective parts, to obtain the following balance 
\begin{equation}\label{eq:balance.microtorques.1}
\intPt \bs{r} (\Div\bs{\xi} + \pi + \gamma ) \dv_\tau + \intdPt \bs{r} (\Divs(\Prj{\bs{n}}\bs{\tau}) + \varpi + \zeta - \xis) \da_\tau = \bs{0}.
\end{equation}
Next, we employ the following identities
\begin{equation}\label{eq:identities.microtorques.balance}
\Div(\bs{r} \otimes \bs{\xi}) = \bs{\xi} + \bs{r} \mskip3mu \Div\bs{\xi}, \qquad \text{and} \qquad \Divs(\bs{r} \otimes \Prj{\bs{n}}\bs{\tau}) = \Prj{\bs{n}}\bs{\tau} + \bs{r} \mskip3mu \Divs(\Prj{\bs{n}}\bs{\tau}),
\end{equation}
followed by the application of the divergence theorem and the divergence theorem on nonsmooth closed surfaces \eqref{eq:nonsmooth.divs.theo.closed.S}, to write expression \eqref{eq:balance.microtorques.1} as 
\begin{equation}\label{eq:balance.microtorques.balance.2}
\intPt ( \bs{r} ( \pi + \gamma ) - \bs{\xi} )\dv_\tau + \intdPt (\bs{r} \otimes \bs{\xi}) \bs{n} \da_\tau +  \intdPt (\bs{r} (\varpi + \zeta - \xis) - \Prj{\bs{n}}\bs{\tau} ) \da_\tau  + \intddPt \surp{(\bs{r} \otimes \bs{\tau}) \bs{\nu}} \ds_\tau = \bs{0}.
\end{equation}
Then, in view of the microtractions \eqref{eq:microtractions}, we arrive at the partwise bulk-surface balance of microtorques
\begin{equation}\label{eq:balance.microtorques.balance.3}
\cl{T}^\flat(\prt_\tau,\dprt_\tau) \coloneqq \intPt ( \bs{r} ( \pi + \gamma ) - \bs{\xi} )\dv_\tau + \intdPt (\bs{r} (\varpi + \zeta ) - \Prj{\bs{n}}\bs{\tau} ) \da_\tau  + \intddPt \bs{r} \tauds \ds_\tau = \bs{0}.
\end{equation}

\section{Conserved species transport}\label{sc:conserved.species.transport}

To account for the transport of species in our system, we supplement the field equations \eqref{eq:field.equations.phase} with the partwise bulk-surface balance of species. We define $\vphiP$ as the bulk mass fraction of one of the species in the bulk, and $\vphidP$ as its surface counterpart. Moreover, since the sum of the mass fractions of both species in the bulk is one, we write $\vphiP$ for one of the bulk mass fraction and $1-\vphiP$ for the other bulk mass fraction. Similarly, on the surface, the surface mass fraction of one species is $\vphidP$ and the other one is $1-\vphidP$.

Next, recallling that $\rhodP = \rho \ell_\tau$, we define the total partwise balance of species as 
\begin{equation}\label{eq:total.partwise.species.balance}
\dot{\overline{\intPt\rhoP \vphiP \dv_\tau}} + \dot{\overline{\intdPt\rho\ell_\tau\vphidP \da_\tau}} = \intPt\sP \dv_\tau + \intdPt \sdP \da_\tau - \intddPt \surp{\bsjdP \cdot \bs{\nu}} \ds_\tau.
\end{equation}
Here, $\sP$ is the bulk species supply, $\sdP$ is the surface species supply, and $\bsjdP$ is the surface species flux.

We also stipulate that the partwise bulk-surface balance of species \eqref{eq:total.partwise.species.balance} may be uncoupled into
\begin{align}\label{eq:bulk.partwise.species.balance}
\dot{\overline{\intPt \rhoP \vphiP \dv_\tau}} &= \intPt \rhoP\dvphiP\dv_\tau, \nonumber\\[4pt]
&= \intPt \sP \dv_\tau - \intdPt \bsjP \cdot \bs{n} \da_\tau,
\end{align}
and
\begin{align}\label{eq:surf.partwise.species.balance}
\dot{\overline{\intdPt \rho \ell_\tau\vphidP \da_\tau}} = {} & \intdPt \rho \ell_\tau\dvphidP\da_\tau, \nonumber \\[4pt]
= {} & \intdPt \mskip3mu \bsjP \cdot \bs{n} \da_\tau + \intdPt \sdP \da_\tau - \intddPt \surp{\bsjdP \cdot \bs{\nu}} \ds_\tau,
\end{align}
where we have used identities \eqref{eq:id.reynolds.mass.P} and \eqref{eq:id.reynolds.mass.dP}.
Here, $\bsjP$ is the bulk species flux.

Next, we apply the divergence and the surface divergence theorems to the uncoupled partwise balance of species \eqref{eq:bulk.partwise.species.balance} and \eqref{eq:surf.partwise.species.balance}. After localization, we are led to the following pointwise bulk balance of species
\begin{equation}\label{eq:pointwise.bulk.mass.balance}
\rhoP\dvphiP = \sP - \Div \bsjP, \qquad \text{in } \prt_\tau,
\end{equation}
and for the surface, using similar arguments which led us to expression \eqref{eq:ell.balance}, we have that the pointwise surface balance of species reads
\begin{equation}\label{eq:pointwise.surface.mass.balance}
\rho\ell_\tau \dvphidP = \bsjP \cdot \bs{n} + \sdP - \Divs(\Prj{\bs{n}}\bsjdP), \qquad \text{on } \dprt_\tau.
\end{equation}
Note that the pointwise bulk balance of species has a standard form, motivating our choice for the uncoupling. However, the pointwise surface balance of mass has a contribution from the bulk, namely $\bsjP \cdot \bs{n}$. Additionally, the term $\Divs (\Prj{\bs{n}}\bsjdP)$ may be split as $\Divs (\Prj{\bs{n}}\bsjdP) = \Divs \bsjdP + 2K \bsjdP \cdot \bs{n}$. Then, the pointwise surface balance of mass \eqref{eq:pointwise.surface.mass.balance} may be written as
\begin{equation}\label{eq:pointwise.surface.mass.balance.alt}
\rho \ell_\tau \dvphidP = \mskip3mu \bsjP \cdot \bs{n} + \sdP - \Divs \bsjdP - 2K \bsjdP \cdot \bs{n}, \qquad \text{on } \dprt_\tau.
\end{equation}

\section{Free-energy imbalance}
\label{sc:free.energy.imbalance}

First, note that the actual external and internal power are given by
\begin{equation}
\left\{
\begin{aligned}
\cl{W}_{\mathrm{ext}}(\prt_\tau,\dprt_\tau) &\coloneqq \cl{V}_{\mathrm{ext}}(\prt_\tau,\dprt_\tau;\bsvP,\dvphiP,\bsvdP,\dvphidP), \quad \text{and} \\[4pt]
\cl{W}_{\mathrm{int}}(\prt_\tau,\dprt_\tau) &\coloneqq \cl{V}_{\mathrm{int}}(\prt_\tau,\dprt_\tau;\bsvP,\dvphiP,\bsvdP,\dvphidP).
\end{aligned}
\right.
\end{equation}
Furthermore, we let $\cl{W}^{\mathrm{conv}}_{\mathrm{ext}}(\prt_\tau,\dprt_\tau)$ denote the  conventional power, which does not include inertial effects, allowing us to write the following relation
\begin{equation}\label{eq:conventional.ext.power}
- \intPt\rhoP\bsvP\cdot\dbsvP\dv_\tau - \intdPt\rho \ell_\tau \bsvdP\cdot\dbsvdP\dv_\tau + \cl{W}_{\mathrm{ext}}^{\mathrm{conv}}(\prt_\tau,\dprt_\tau) = \cl{W}_{\mathrm{ext}}(\prt_\tau,\dprt_\tau) = \cl{W}_{\mathrm{int}}(\prt_\tau,\dprt_\tau),
\end{equation}
where expressions \eqref{eq:generalised.body.force} and \eqref{eq:inertial.body.force} were used.

In the free-energy imbalance, we account for the rate at which energy is transferred to $\prt_\tau$ and $\dprt_\tau$ due to species transport, as well as the external power expenditure. Thus, the free-energy imbalance reads
\begin{align}\label{eq:partwise.free.energy.imbalance.gen}
\dot{\overline{\intPt\rhoP(\psiP+\fr{1}{2}|\bsvP|^2)\dv_\tau}} + \dot{\overline{\intdPt\rho\ell_\tau(\psidP+\fr{1}{2}|\bsvdP|^2)\da_\tau}} \le {}& \cl{W}^{\mathrm{conv}}_{\mathrm{ext}}(\prt_\tau,\dprt_\tau) \nonumber\\[4pt]
&+ \intPt \muP \sP \dv_\tau - \intdPt \muP \bsjP \cdot \bs{n} \da_\tau \nonumber\\[4pt]
&+ \intdPt \mudP \bsjP \cdot \bs{n} \da_\tau + \intdPt \mudP \sdP \da_\tau \nonumber\\[4pt]
&- \intddPt \surp{\mudP \bsjdP \cdot \bs{\nu}} \ds_\tau.
\end{align}
Here, $\psiP$ is the bulk free-energy density, $\psidP$ is the surface free-energy density, $\muP$ is the bulk chemical potential, and $\mudP$ is the surface chemical potential. Next, using \eqref{eq:conventional.ext.power} in \eqref{eq:partwise.free.energy.imbalance.gen} along with expressions \eqref{eq:id.reynolds.mass.P} and \eqref{eq:id.reynolds.mass.dP}, we arrive at
\begin{align}\label{eq:partwise.free.energy.imbalance}
\intPt\rhoP\dpsiP\dv_\tau + \intdPt\rho\ell_\tau\dpsidP\da_\tau \le {}& \cl{W}_{\mathrm{int}}(\prt_\tau,\dprt_\tau) \nonumber\\[4pt]
&+ \intPt \muP \sP \dv_\tau - \intdPt \muP \bsjP \cdot \bs{n} \da_\tau \nonumber\\[4pt]
&+ \intdPt \mudP \bsjP \cdot \bs{n} \da_\tau + \intdPt \mudP \sdP \da_\tau \nonumber\\[4pt]
&- \intddPt \surp{\mudP \bsjdP \cdot \bs{\nu}} \ds_\tau.
\end{align}

Noting that $\Prj{\bs{n}}\bsjdP \cdot \Grads\mudP = \bsjdP \cdot \Grads\mudP$ with expressions \eqref{eq:generalised.body.force}, \eqref{eq:pointwise.bulk.mass.balance}, and \eqref{eq:pointwise.surface.mass.balance}, and using the divergence theorem and the surface divergence theorem for nonsmooth closed surfaces \eqref{eq:nonsmooth.divs.theo.closed.S}, we arrive at
\begin{align}\label{eq:partwise.free.energy.imbalance.final}
\intPt\rhoP\dpsiP\dv_\tau + \intdPt\rho\ell_\tau\dpsidP\da_\tau \le {}& \intPt \br{T} \twovdots \Grad \bsvP \dv_\tau + \intdPt \br{H} \twovdots \Grads\bsvdP \da_\tau \nonumber\\[4pt]
&+ \intPt \bs{\xi} \cdot \Grad \dvphiP \dv_\tau - \intPt \pi \dvphiP \dv_\tau \nonumber\\[4pt]
&+ \intdPt \bs{\tau} \cdot \Grads \dvphidP \da_\tau - \intdPt \varpi \dvphidP \da_\tau \nonumber\\[4pt]
&+ \intPt \muP (\sP - \Div\bsjP) \dv_\tau - \intPt \bsjP \cdot \Grad \muP \dv_\tau \nonumber\\[4pt]
&+ \intdPt \mudP (\bsjP \cdot \bs{n} + \sdP - \Divs \bsjdP) \da_\tau - \intdPt \bsjdP \cdot \Grads \mudP \da_\tau.
\end{align}
Next, using the balance of species in bulk \eqref{eq:pointwise.bulk.mass.balance} and on the surface \eqref{eq:pointwise.surface.mass.balance}, we obtain the uncoupled pointwise imbalances, which satisfy the partwise bulk-surface free-energy imbalance \eqref{eq:partwise.free.energy.imbalance.final}. Thus, the pointwise bulk free-energy imbalance and the pointwise surface free-energy imbalance, respectively, read
\begin{equation}
\rhoP\dpsiP - \br{T}\twovdots\Grad\bsvP + (\pi - \rhoP\muP) \dvphiP - \bs{\xi} \cdot \Grad\dvphiP + \bsjP \cdot \Grad\muP \le 0, \quad \text{in } \prt_\tau,
\end{equation}
and
\begin{equation}\label{eq:pointwise.surface.free.energy.localization}
\rho\ell_\tau \dpsidP - \br{H}\twovdots\Grads\bsvdP + (\varpi - \rho\ell_\tau\mudP) \dvphidP - \bs{\tau} \cdot \Grads\dvphidP + \bsjdP \cdot \Grads\mudP \le 0, \quad \text{on } \dprt_\tau,
\end{equation}
where, to arrive at expression \eqref{eq:pointwise.surface.free.energy.localization}, we have used similar arguments which led us to expression \eqref{eq:ell.balance}. Using the following identities
\begin{equation}
\Grad\dvphiP = (\Grad\vphiP)\dot{\vphantom{\varphi}} + (\Grad\bsvP)^{\trans} \Grad\vphiP,
\end{equation}
and
\begin{equation}
\Grads\dvphidP = (\Grads\vphidP)\mathring{\vphantom{\varphi}} + (\Grads\bsvdP)^{\trans} \Grads\vphidP,\footnotemark
\end{equation}
\footnotetext{This identity may be obtained from the previous one by premultiplying by $\Prj{n}$ and assuming a normal constant extension of $\bsvdP$ and $\vphidP$ while noting that $\partial_n \vphidP = 0$.}the free-energy imbalances become
\begin{equation}\label{eq:general.bulk.free.energy.imbalance}
\rhoP\dpsiP - (\br{T} + \Grad\vphiP \otimes \bs{\xi})\twovdots\Grad\bsvP + (\pi - \rhoP\muP) \dvphiP - \bs{\xi} \cdot (\Grad\vphiP)\dot{\vphantom{\varphi}} + \bsjP \cdot \Grad\muP \le 0, \quad \text{in } \prt_\tau,
\end{equation}
and
\begin{multline}\label{eq:general.surface.free.energy.imbalance}
\rho\ell_\tau \dpsidP - (\br{H} + \Grads\vphidP \otimes \bs{\tau})\twovdots\Grads\bsvdP + (\varpi - \rho\ell_\tau \mudP) \dvphidP \\[4pt] - \bs{\tau} \cdot (\Grads\vphidP)\mathring{\vphantom{\varphi}} + \bsjdP \cdot \Grads\mudP \le 0, \quad \text{on } \dprt_\tau.
\end{multline}

\section{Constitutive response functions}
\label{sc:constitutive.relations}

The set of independent and dependent variables are, respectively, given by $\{\rhodP$, $\vphiP$, $\vphidP$, $\Grad\vphiP$, $\Grads\vphidP$, $\muP$, $\mudP$, $\Grad \bsvP$, $\Grads \bsvdP\}$ and $\{\pi$, $\varpi$, $\bs{\xi}$, $\bs{\tau}$, $\bsjP$, $\bsjdP$, $\brT$, $\brH\}$. Thus, we find that the local inequalities \eqref{eq:general.bulk.free.energy.imbalance} and \eqref{eq:general.surface.free.energy.imbalance} are satisfied in all processes if:
\begin{itemize}
\item The bulk and surface free-energy densities $\psiP$ and $\psidP$ are, respectively, given by constitutive response functions that are independent of $\muP$, $\mudP$, $\Grad\muP$, and $\Grads\mudP$:
\begin{equation}
\psiP \coloneqq \psiP(\vphiP,\Grad\vphiP), \qquad \text{and} \qquad \psidP \coloneqq \psidP(\rhodP,\vphidP,\Grads\vphidP).
\end{equation}
Thus, with
\begin{equation}
\dpsiP = \pvphiP \psiP \dvphiP + \pgvphiP \psiP \cdot (\Grad\vphiP)^{\bs{\dot{}}},
\end{equation}
and
\begin{align}
\dpsidP &= \prhodP \psidP \drhodP + \pvphidP \psidP \dvphidP + \pgvphidP \psidP \cdot (\Grads\vphidP)\mathring{\vphantom{\varphi}}, \nonumber\\[4pt]
{} &= -\ell_\tau \pelldP \psidP \mskip3mu \Divs \bsvdP + \pvphidP \psidP \dvphidP + \pgvphidP \psidP \cdot (\Grads\vphidP)\mathring{\vphantom{\varphi}}, 
\end{align}
we are led to two pointwise free-energy imbalances
\begin{multline}\label{eq:bulk.free.energy.imbalance}
(\br{T} + \Grad\vphiP \otimes \bs{\xi})\twovdots\Grad\bsvP + (\rhoP \muP - \pi - \rhoP \pvphiP \psiP) \dvphiP \\[4pt]
+ (\bs{\xi} - \rhoP \pgvphiP \psiP) \cdot (\Grad\vphiP)\dot{\vphantom{\varphi}} - \bsjP \cdot \Grad\muP \ge 0, \quad \text{in } \prt_\tau,
\end{multline}
and
\begin{multline}\label{eq:surface.free.energy.imbalance}
(\br{H} + \Grads\vphidP \otimes \bs{\tau})\twovdots\Grads\bsvdP + \rho \ell_\tau^2 \pelldP \psidP \mskip3mu \Divs \bsvdP + (\rho\ell_\tau\mudP - \varpi - \rho\ell_\tau\pvphidP \psidP) \dvphidP \\[4pt]
+ (\bs{\tau} - \rho\ell_\tau\pgvphidP \psidP) \cdot (\Grads\vphidP)\mathring{\vphantom{\varphi}} - \bsjdP \cdot \Grads\mudP \ge 0, \quad \text{on } \dprt_\tau.
\end{multline}
From a microscopic standpoint, one expects to have a constant microscopic free-energy density for a fixed $\vphidP$ and $\Grads\vphidP$ in the thin film of fluid surrounding the bulk fluid. However, since the thickness may change, one may stipulate that the surface free-energy density, therefore, a macroscopic quantity, depends linearly on thickness $\ell_\tau$. That is, we may define the surface free-energy density as
\begin{equation}
\psidP(\rhodP,\vphidP,\Grads\vphidP) \coloneqq \dfrac{\rhodP}{\rho} \psi(\vphidP,\Grads\vphidP) = \ell_\tau \psi(\vphidP,\Grads\vphidP),
\end{equation}
where $\psi$ is the microscopic free-energy density in the thin film of the fluid. Therefore, the corresponding pointwise free-energy imbalance is given by 
\begin{multline}\label{eq:surface.free.energy.imbalance.restricted}
(\br{H} + \Grads\vphidP \otimes \bs{\tau})\twovdots\Grads\bsvdP + \rho \ell_\tau^2 \psi \mskip3mu \Divs \bsvdP + (\rho\ell_\tau\mudP - \varpi - \rho\ell_\tau^2 \pvphidP \psi) \dvphidP \\[4pt]
+ (\bs{\tau} - \rho\ell_\tau^2\pgvphidP \psi) \cdot \Grads\dvphidP - \bsjdP \cdot \Grads\mudP \ge 0 \quad \text{on } \dprt_\tau.
\end{multline}
\item The bulk and surface microstress $\bs{\xi}$ and $\bs{\tau}$ are, respectively, given by
\begin{equation}\label{eq:microstress}
\bs{\xi} \coloneqq \rhoP\pgvphiP \psiP, \qquad \text{and} \qquad \bs{\tau} \coloneqq \rho\ell_\tau \pgvphidP \psidP.
\end{equation}
\item The internal bulk and surface microforces $\pi$ and $\varpi$ are, respectively, given by constitutive response functions that differ from the bulk and surface chemical potential by a contribution derived, respectively, from the response functions $\psiP$ and $\psidP$
\begin{equation}\label{eq:internal.microforces}
\pi \coloneqq \rhoP(\muP - \pvphiP \psiP), \qquad \text{and} \qquad \varpi \coloneqq \rho\ell_\tau(\mudP - \pvphidP \psidP).
\end{equation}
\item Granted that the bulk and surface species fluxes $\bsjP$ and $\bsjdP$ depend smoothly on the gradient of the bulk chemical potential, $\Grad\muP$, and the surface gradient of the surface chemical potential, $\Grads\mudP$, these fluxes are, respectively, given by a constitutive response function of the form
\begin{equation}\label{eq:fluxes}
\bsjP \coloneqq - \bsMP \Grad\muP, \qquad \text{and} \qquad \bsjdP \coloneqq -\bsMdP \Grads\mudP,
\end{equation}
where the mobility tensors $\bsMP$ and $\bsMdP$ must, respectively, obey the residual dissipation inequalities
\begin{equation}\label{eq:res.diss.inequalities}
\Grad\muP \cdot \bsMP \Grad\muP \ge 0, \qquad \text{and} \qquad \Grads\mudP \cdot \bsMdP \Grads\mudP \ge 0.
\end{equation}
Oftentimes one finds in the literature that $\bsMP \coloneqq \mP \id$ and $\bsMdP \coloneqq \mdP \Prj{\bs{n}}$, where $\mP$ and $\mdP$ denote the scalar bulk and surface mobilities, respectively. With this choice for $\bsMdP$, the surface flux $\bsjdP$ remains proportional to $\Grads\mudP$ and therefore tangential to $\dprt_\tau$. As a result of the vanishing contributions of $\bsMdP$ to the normal component of $\bsjdP$, the pointwise surface species balance \eqref{eq:pointwise.surface.mass.balance.alt} becomes
\begin{equation}\label{eq:pointwise.surface.mass.balance.alt.restricted}
\rho \ell_\tau \dvphidP =  \bsjP \cdot \bs{n} + \sdP - \Divs \bsjdP, \qquad \text{on } \dprt_\tau.
\end{equation}
\item The bulk and surface stress are, respectively, given by
\begin{equation}\label{eq:stresses}
\left\{
\begin{aligned}
\br{T} &\coloneqq \brTvis(\DP) - \pPmech \id - \Grad\vphiP \otimes \bs{\xi}, \\[4pt]
\br{H} &\coloneqq \brHvis(\rhodP, \DdP) - \pdPtherm \Prj{\bs{n}} - \Grads\vphidP \otimes \bs{\tau},
\end{aligned}
\right.
\end{equation}
where the stretching tensor is defined as $\DP \coloneqq \sym \mskip2mu \Grad \bsvP$ and $\DdP \coloneqq \sym \mskip2mu \Grads \bsvdP$, with $\sym$ the symmetric operator. Furthermore, $\pPmech$ is the mechanical bulk pressure. In addition, $\pdPtherm$ denotes the thermodynamical surface pressure, which is oftentimes called surface tension. These pressures are discussed in more detail in what follows. Additionally, we consider that the viscous bulk and surface stress $\brTvis$ and $\brHvis$ are, respectively, linear in $\DP$ and $\DdP$, that is,
\begin{equation}\label{eq:viscous.stresses}
\left\{
\begin{aligned}
\br{T}^{\mathrm{vis}} &\coloneqq 2\bmuP \mskip3mu \DP, \\[4pt]
\br{H}^{\mathrm{vis}} &\coloneqq 2\bmudP \mskip3mu \Prj{\bs{n}} \DdP \Prj{\bs{n}} + (\barkappadP - \bmudP) (\Divs \bsvdP)\Prj{\bs{n}}, \\[4pt]
&= 2\bmudP \mskip3mu \Prj{\bs{n}} \DdP \Prj{\bs{n}} - \dfrac{\mathring{\ell}_\tau}{\ell_\tau} (\barkappadP - \bmudP) \Prj{\bs{n}},
\end{aligned}
\right.
\end{equation}
where $\bmuP$ and $\bmudP(\rhodP)$ denote the bulk and surface dynamic viscosities\footnote{$\ \bmuP$ and $\bmudP(\rhodP)$ are also referred to as the bulk and surface shear viscosities, respectively.}, and $\ \barkappadP(\rhodP)$ is the surface dilatational viscosity\footnote{$\barkappadP$ is also referred to as the bulk viscosity of the surface fluid.}. Also, note that, in \eqref{eq:viscous.stresses}$_2$, we have used the surface mass balance \eqref{eq:ell.balance}.
Next, we decompose $\DP$ and $\DdP$ into a deviatoric and spherical part, that is, $\DP = \DkP + \fr{1}{3}(\tr\DP)\id$ and $\DdP = \DkdP + \fr{1}{2}(\tr\DdP)\Prj{\bs{n}}$, with $\tr(\DkP) = \tr(\DkdP) = 0$. Then, we may compute the viscous dissipation as
\begin{equation}\label{eq:viscous.dissipation}
\left\{
\begin{aligned}
\br{T}^{\mathrm{vis}} \twovdots \DP &= \Big(2\bmuP \mskip3mu \DP\Big) \twovdots \Big(\DkP + \fr{1}{3}(\tr\DP) \id \Big), \\[4pt]
&= 2\bmuP |\DkP|^2, \\[4pt]
\br{H}^{\mathrm{vis}} \twovdots \DdP &= \Big(2\bmudP \mskip3mu \Prj{\bs{n}} \DdP \Prj{\bs{n}} + (\barkappadP - \bmudP)(\Divs \bsvdP)\Prj{\bs{n}}\Big) \twovdots \Big(\DkdP + \fr{1}{2}(\tr\DdP) \Prj{\bs{n}}\Big), \\[4pt]
&=2\bmudP |\Prj{\bs{n}}\DkdP|^2 + \barkappadP(\Divs \bsvdP)^2, \\[4pt]
&=2\bmudP |\Prj{\bs{n}}\DkdP|^2 + \barkappadP \Bigg(\dfrac{\mathring{\ell}_\tau}{\ell_\tau}\Bigg)^2,
\end{aligned}
\right.
\end{equation}
where we have taken into account that $\tr(\DP) = \Div \bsvP = 0$ and $\tr(\DdP) = \Divs\bsvdP = -\fr{\mathring{\ell}_{\tau}}{\ell_{\tau}}$.
In view of the free-energy inequalities in \eqref{eq:bulk.free.energy.imbalance} and \eqref{eq:surface.free.energy.imbalance}, we conclude that $\bmuP \geq 0$, $\bmudP(\rho \ell_\tau) \geq 0$ and $ \barkappadP(\rho \ell_\tau)\geq 0$.
\item The total pressure in the bulk and surface fluids are, respectively, given by 
\begin{equation}\label{eq:total.pressure}
\left\{
\begin{aligned}    
& - \fr{1}{3} \tr{(\brT)} = \pPmech + \pPcap \eqqcolon \pPtot,\\[4pt]
& - \fr{1}{2} \tr{(\brH)} = \pdPtherm + \pdPmech + \pdPcap \eqqcolon \pdPtot.
\end{aligned} \right.
\end{equation}
In the above, the mechanical bulk pressure $\pP$ is indeterminate, whereas the bulk capillary-like pressure is defined as $\pPcap \coloneqq- \fr{1}{3}\tr{(\Grad\vphiP \otimes \bs{\xi})}$. For the surface, we find $\pdPcap \coloneqq - \fr{1}{2}\tr{(\Grads\vphidP \otimes \bs{\tau})}$. In view of the surface free-energy imbalance \eqref{eq:surface.free.energy.imbalance.restricted} and the viscous surface dissipation \eqref{eq:viscous.dissipation}$_2$, the thermodynamical surface pressure is given by 
\begin{equation}\label{eq.thermodynamical.pressure}
\pdPtherm \coloneqq \rho\ell_\tau^2\psi, 
\end{equation}
and, lastly, $\pdPmech \coloneqq - \fr{1}{2} \tr(\br{H}^{\mathrm{vis}}) = - \barkappadP \mskip3mu \Divs \bsvdP$. Thus, the total bulk and surface pressure read
\begin{equation}\label{eq:total.pressure.revised} 
\left\{
\begin{aligned}    
\pPtot &\coloneqq \pPmech - \fr{1}{3}\tr{(\Grad\vphiP \otimes \bs{\xi})},\\[4pt]
\pdPtot &\coloneqq \rho\ell_\tau^2\psi - \barkappadP \mskip3mu \Divs \bsvdP - \fr{1}{2}\tr{(\Grads\vphidP \otimes \bs{\tau})}, \\[4pt]
&= \rho\ell_\tau^2\psi + \barkappadP \dfrac{\mathring{\ell}_\tau}{\ell_\tau} - \fr{1}{2}\tr{(\Grads\vphidP \otimes \bs{\tau})}.
\end{aligned} \right.
\end{equation}
\item In view of \eqref{eq:viscous.stresses} and \eqref{eq.thermodynamical.pressure}, we conclude that stresses in \eqref{eq:stresses} can be rewritten as
\begin{equation}\label{eq:stresses.revised}
\left\{
\begin{aligned}
\br{T} &\coloneqq 2\bmuP \mskip3mu \DP - \pPmech \id - \Grad\vphiP \otimes \bs{\xi}, \qquad \text{and} \qquad \\[4pt]
\br{H} &\coloneqq 2\bmudP \mskip3mu  \Prj{\bs{n}} \DkdP  \Prj{\bs{n}} - (\rho\ell_\tau^2\psi - \barkappadP \Divs \bsvdP)\Prj{\bs{n}} - \Grads\vphidP \otimes \bs{\tau}, \\[4pt]
&= 2\bmudP \mskip3mu \Prj{\bs{n}} \DkdP  \Prj{\bs{n}}  - \Bigg(\rho\ell_\tau^2\psi + \barkappadP\dfrac{\mathring{\ell}_\tau}{\ell_\tau}\Bigg)\Prj{\bs{n}} - \Grads\vphidP \otimes \bs{\tau},
\end{aligned}
\right.
\end{equation}
Note that the bulk and surface stress in \eqref{eq:stresses.revised} are symmetric, that is, $\br{T} = \br{T}^{\trans}$ and $\br{H} = \br{H}^{\trans}$. Due to the frame-indifference result in \eqref{eq:cond.H}, surface stresses must also annihilate the normal, that is, $\br{H}\bs{n} = \bs{0}$, which is also satisfied by \eqref{eq:stresses.revised}$_2$. Additionally, the bulk and surface capillary contributions to the bulk and surface stress, namely $\Grad\vphiP \otimes \bs{\xi}$ and $\Grads\vphidP \otimes \bs{\tau}$, respectively, must also be symmetric. This implies that the expressions for the bulk and surface microstress given in \eqref{eq:microstress} must be, respectively, proportional to $\Grad\vphiP$ and $\Grads\vphidP$.
\end{itemize}

Lastly, we use the above results to present the explicit form of the bulk and surface chemical potentials. Substitution of \eqref{eq:microtractions} and \eqref{eq:field.equations.phase} in \eqref{eq:internal.microforces} yields
\begin{equation}
\rhoP \muP = - \Div\bs{\xi} - \gamma + \rhoP \pvphiP \psiP, \qquad \text{in } \prt_\tau,
\end{equation}
and
\begin{equation}
\rho\ell_\tau \mudP = - \Divs(\Prj{\bs{n}}\bs{\tau}) - \zeta + \bs{\xi} \cdot \bs{n} + \rho \ell_\tau \pvphidP \psidP, \qquad \text{on } \dprt_\tau.
\end{equation}
In view of \eqref{eq:microstress}, these expressions take the following form 
\begin{equation}
\rhoP \muP = \rhoP \pvphiP \psiP - \Div\Big(\rhoP \pgvphiP \psiP\Big) - \gamma, \qquad \text{in } \prt_\tau,
\end{equation}
and, recalling that $\rhodP = \rho \ell_\tau$, we obtain
\begin{equation}\label{eq:surface.chemical.potential}
\rho \ell_\tau \mudP = \rho \ell_\tau^2 \pvphidP \psi - \Divs\Big(\rho\ell_\tau^2 \pgvphidP \psi \Big) - \zeta + \rhoP \pgvphiP \psiP \cdot \bs{n}, \qquad \text{on } \dprt_\tau.
\end{equation}
Additionally, the term $\Divs (\rho \ell_\tau^2 \pgvphidP \psi)$ may be split as $\Divs (\rho \ell_\tau^2 \pgvphidP \psi) = \rho \ell_\tau^2 \Divs( \pgvphidP \psi)$ $+ 2 \rho \ell_\tau \pgvphidP \psi \cdot \Grads \ell_\tau $. Then, expression \eqref{eq:surface.chemical.potential} can be rewritten as
\begin{equation}
\rho \ell_\tau \mudP = \rho \ell_\tau^2 \pvphidP \psi - \rho \ell_\tau^2 \Divs( \pgvphidP \psi) - 2\rho \ell_\tau \pgvphidP \psi \cdot \Grads \ell_\tau - \zeta + \rhoP \pgvphiP \psiP \cdot \bs{n}, \quad \text{on } \dprt_\tau.
\end{equation}

\section{Specialized equations}
\label{sc:specialization}

To exemplify our bulk-surface continuum theory for fluid flow undergoing phase segregation, we consider the following free-energy densities
\begin{equation} \label{eq:free.energy.densities.specialized}
\psiP \coloneqq \fr{1}{\epsilon} f(\vphiP) + \fr{\epsilon}{2} |\Grad\vphiP|^2, \qquad \text{and} \qquad \psidP \coloneqq \ell_\tau \Big( \fr{1}{\delta} g(\vphidP) + \fr{\iota \delta}{2} |\Grads\vphidP|^2 \Big),
\end{equation}
so that the total free-energy functional reads
\begin{multline}\label{eq:free.energy.functional.specialized}
\intPt \rhoP \psiP \dv_\tau + \intdPt \rho \ell_\tau \psidP \da_\tau \\[4pt]
= \intPt \rhoP \Big( \fr{1}{\epsilon} f(\vphiP) + \fr{\epsilon}{2} |\Grad\vphiP|^2 \Big) \dv_\tau + \intdPt \rho \ell_\tau^2 \Big( \fr{1}{\delta} g(\vphidP) + \fr{\iota \delta}{2} |\Grads\vphidP|^2 \Big) \da_\tau,
\end{multline}
which describes the phase segregation in our bulk-surface system. Here, $\epsilon$, $\delta$, and $\iota$ are real positive constant parameters, while $f$ and $g$ denote the bulk and surface potentials, respectively. With this choice for the free-energy densities, the bulk and surface microstress \eqref{eq:microstress} specialize to
\begin{equation}\label{eq:microstress.specialized}
\bs{\xi} \coloneqq \epsilon \rhoP \Grad \vphiP, \qquad \text{and} \qquad 
\bs{\tau} \coloneqq \iota \delta \rho \ell_\tau^2 \Grads\vphidP, 
\end{equation}
while the internal bulk and surface microforce \eqref{eq:internal.microforces} become
\begin{equation}\label{eq:internal.microforces.specialized}
\pi \coloneqq \rhoP(\muP - \fr{1}{\epsilon} f'(\vphiP)), \qquad \text{and} \qquad 
\varpi \coloneqq \rho\ell_\tau(\mudP - \ell_\tau\fr{1}{\delta}g'(\vphidP)).
\end{equation}
Furthermore, we consider the following bulk and surface species fluxes 
\begin{equation}\label{eq:fluxes.specialized}
\bsjP \coloneqq - \bsMP \Grad\muP, \qquad \text{and} \qquad \bsjdP \coloneqq -\bsMdP \Grads\mudP,
\end{equation}
where, for the sake of simplicity, we use $\bsMP\coloneqq \mP \boldsymbol{I}$ and $\bsMdP \coloneqq \mdP \Prj{\bs{n}}$ with scalar functions $\mP, \mdP \geq 0$ to ensure that bulk and surface mobilities satisfy the residual dissipation inequalities \eqref{eq:res.diss.inequalities}.

Thus, in view of expressions \eqref{eq:microstress.specialized} and \eqref{eq:fluxes.specialized}, the bulk field equation \eqref{eq:field.equations.phase}$_1$ takes on the following form
\begin{equation}\label{eq:species.bulk.spezialized}
\rhoP \muP = \fr{1}{\epsilon} \rhoP f^\prime(\vphiP) - \epsilon \rhoP \triangle\vphiP - \gamma, \qquad \text{in } \prt_\tau,
\end{equation}
where $\triangle \coloneqq \Div \Grad$ denotes the Laplace operator. Additionally, for the above choices the surface species equation  \eqref{eq:field.equations.fluid.mod}$_2$ can be written as
\begin{equation}\label{eq:species.surface.spezialized}
\rho \ell_\tau \mudP = \fr{1}{\delta} \rho \ell_\tau^2 g^\prime(\vphidP)
- \iota \delta \rho \ell_\tau (\ell_\tau \triangles \vphidP
+ 2 \mskip3mu \Grads \vphidP \cdot \Grads \ell_\tau)
+ \epsilon \rhoP \mskip3mu \Grad\vphiP \cdot \boldsymbol{n} - \zeta,  \qquad \text{on } \dprt_\tau,
\end{equation}
with $\triangles \coloneqq \Divs \Grads$ denoting the Laplace--Beltrami operator, see also definition \eqref{eq:laplace.beltrami}. Note that we have arrived at equation \eqref{eq:species.surface.spezialized} using the surface microtraction \eqref{eq:microtractions}$_1$, as well as $\Grads \vphidP \cdot \boldsymbol{n} = 0$.

Next, we derive the specialized equations of motion. First, using the surface free-energy density \eqref{eq:free.energy.densities.specialized}$_2$ and the microstresses \eqref{eq:microstress.specialized}, the bulk and surface stresses \eqref{eq:stresses.revised} specialize to
\begin{equation} \label{eq:stresses.specialized}
\left\{
\begin{aligned}
\br{T} &\coloneqq 2\bmuP \mskip3mu \DP - \pPmech \id - \epsilon \rhoP \Grad\vphiP \otimes \Grad\vphiP, \qquad \text{and} \qquad \\[4pt]
\br{H} &\coloneqq 2\bmudP \mskip3mu  \Prj{\bs{n}} \DkdP  \Prj{\bs{n}}  - \Bigg(\rho\ell_\tau^2 \Big( \fr{1}{\delta} g(\vphidP) + \fr{\iota \delta}{2} |\Grads\vphidP|^2 \Big)  + \barkappadP\dfrac{\mathring{\ell}_\tau}{\ell_\tau}\Bigg)\Prj{\bs{n}} \\[4pt]
&\quad - \iota \delta \rho\ell_\tau \mskip3mu \Grads\vphidP \otimes \Grads\vphidP,
\end{aligned}
\right.
\end{equation}
recalling that $\DP \coloneqq \sym \mskip2mu \Grad \bsvP$, $\DdP \coloneqq \sym \mskip2mu \Grads \bsvdP$ and that $\DkdP$ is the deviatoric part of $\DdP$.
Using these bulk stress \eqref{eq:stresses.specialized}$_1$, the bulk equation of motion \eqref{eq:field.equations.fluid.explicit}$_1$ takes on the form 
\begin{align}\label{eq:motion.bulk.spezialized}
\rhoP \dbsvP ={}& \Div\br{T} + \bsbni, \nonumber\\[4pt]
={}& 2 \mskip3mu \Div (\bmuP \mskip3mu\DP) - \Grad \pPmech - \epsilon \rhoP \mskip3mu \Div (\Grad\vphiP \otimes \Grad\vphiP) + \bsbni, \qquad \text{in } \prt_\tau.
\end{align}
Conversely, we arrive at the specialized surface equation of motion by using the surface traction \eqref{eq:tractions}$_1$ into surface equation \eqref{eq:field.equations.fluid.explicit}$_2$ and splitting the surface divergence operator as done in \eqref{eq:field.equations.fluid.mod}, followed by substitution of the stresses \eqref{eq:stresses.specialized}, yielding
\begin{alignat}{2}\label{eq:motion.surface.spezialized}
\rho \ell_\tau \dbsvdP  ={}& \Divs\br{H} + 2K \br{H} \bs{n} + \bsgni - \br{T} \bs{n}, \nonumber\\[4pt]
={}& 2 \mskip3mu \Divs(\bmudP \mskip3mu \Prj{\bs{n}} \DkdP  \Prj{\bs{n}})\nonumber - \rho \ell_\tau^2 \mskip3mu \Grads \Big( \fr{1}{\delta} g(\vphidP) + \fr{\iota \delta}{2} |\Grads\vphidP|^2 \Big ) \\[4pt]
&-2 \rho \ell_\tau \Big( \fr{1}{\delta} g(\vphidP) + \fr{\iota \delta}{2} |\Grads\vphidP|^2 \Big) \Grads \ell_\tau - \Grads \Big(\barkappadP\dfrac{\mathring{\ell}_\tau}{\ell_\tau} \Big) \nonumber\\[4pt]
&- \iota \delta \rho \ell_\tau \mskip3mu \Divs \Big( \Grads\vphidP \otimes \Grads\vphidP\Big) - \iota \delta \rho \mskip3mu \Grads \ell_\tau \cdot \Big( \Grads\vphidP \otimes \Grads\vphidP\Big) \nonumber\\[4pt]
&+ \bsgni - 2\bmuP \mskip3mu \DP\bs{n} +  \pPmech \bs{n} + \epsilon \rhoP \mskip3mu \Grad\vphiP \otimes \Grad\vphiP \bs{n}, \qquad \text{on } \dprt_\tau,
\end{alignat}
where we have additionally used that $\Prj{\bs{n}}\bs{n}=\bs{0}$. 

Lastly, supplementing the specialized field equations \eqref{eq:species.bulk.spezialized}, \eqref{eq:species.surface.spezialized}, \eqref{eq:motion.bulk.spezialized}, and \eqref{eq:motion.surface.spezialized} with the isochoric constraints in \eqref{eq:isochoric.constraint.bulk} and \eqref{eq:ell.balance} renders a system of equations that describes the dynamics of the bulk-surface system in terms of the state variables $\vphiP$, $\vphidP$, $\bsvP$, $\bsvdP$, $\pPmech$ and $\ell_\tau$.

\section{Dynamic boundary conditions}
\label{sc:boundary}

We use \emph{dynamic boundary} to refer to a boundary that is endowed with its own mechanical laws, whereas \emph{static boundary} is reserved for a boundary that is arbitrarily prescribed as an function of space and time, yet does not obey any underlying partial differential equation. Therefore, we impose static boundary conditions on $\ddprt_\tau$, as we have not endowed edges with their own mechanical laws. Furthermore, for the sake of simplicity, we only impose dynamic conditions on $\dprt_\tau$, and do not consider any static conditions $\dprt_\tau$, as was done in \cite{Esp23}.

Prior to establishing dynamic boundary conditions on $\dprt_\tau$, let us introduce an environmental surface imbalance. To this end, we here use the partwise surface imbalance presented by Espath \cite{Esp23} based on arguments described by Fried \& Gurtin in \cite[surface free-energy imbalance (92)]{Fri06} on a migrating boundary $\dprt_\tau$. Specifically, we stipulate that 
\begin{equation}\label{eq:free.energy.environment}
\cl{T}_{\mathrm{surf}}(-\dprt_\tau) + \cl{T}_{\mathrm{env}}(\dprt_\tau) \ge 0,
\end{equation}
where $\cl{T}_{\mathrm{surf}}(-\dprt_\tau)$ represents the power expended on $\dprt_\tau$ by the material inside $\prt_\tau$ and $\dprt_\tau$ in addition to the rate at which energy is transferred from $\prt_\tau$ to $\dprt_\tau$. Futhermore, $\cl{T}_{\mathrm{env}}(\dprt_\tau)$ combines the power expended by the environment on $\dprt_\tau$ and the rate at which energy is transferred from the environment to $\dprt_\tau$. Therefore, on $\dprt_\tau$, we define
\begin{align}\label{eq:power.surface.dyn}
\cl{T}_{\mathrm{surf}}(-\dprt_\tau) \coloneqq {}& - \intdPt (\bsvP - \bsvdP) \cdot \bsts \da_\tau - \intddPt \bshds \cdot \bsvdP \ds_\tau \nonumber \\[4pt]
& - \intdPt (\dvphiP - \dvphidP) \xis \da_\tau - \intddPt \tauds \dvphidP \ds_\tau \nonumber\\[4pt]
& - \intdPt (\mudP - \muP) \bsjP \cdot \bs{n} \da_\tau + \intddPt \surp{\mudP \bsjdP \cdot \bs{\nu}} \ds_\tau,
\end{align}
where $\cl{T}_{\mathrm{surf}}(-\dprt_\tau) = -\cl{T}_{\mathrm{surf}}(\dprt_\tau)$.
We define the contribution from the environment as follows
\begin{equation}\label{eq:power.env.dyn}
\cl{T}_{\mathrm{env}}(\dprt_{\tau}) \coloneqq \intddPt \bshds^{\mathrm{env}} \cdot \bsvddP^{\mathrm{env}} \ds_\tau + \intddPt \tauds^{\mathrm{env}} \dvphiddP^{\mathrm{env}} \ds_\tau - \intddPt \muddP^{\mathrm{env}} \jmath_{\scriptscriptstyle\ddprt}^{\mathrm{env}} \ds_\tau.
\end{equation}
Now, on $\ddprt_\tau$, we set $\bshds^{\mathrm{env}}= \bshds$, $\bsvddP^{\mathrm{env}} = \bsvdP$, $\tauds^{\mathrm{env}} = \tauds$, $\dvphiddP^{\mathrm{env}} = \dvphidP$, $\muddP^{\mathrm{env}} = \mudP$, and $\jmath_{\scriptscriptstyle\ddprt}^{\mathrm{env}} = \surp{\bsjdP \cdot \bs{\nu}}$. Thus, the surface free-energy imbalance in \eqref{eq:free.energy.environment} reads
\begin{equation}\label{eq:free.energy.environment.mix}
- \intdPt \big( (\bsvP - \bsvdP) \cdot \bsts + (\dvphiP - \dvphidP) \mskip3mu \xis + (\mudP - \muP) \mskip3mu \bsjP \cdot \bs{n} \big) \da_\tau \ge 0.
\end{equation}
Uncoupling this expression gives us
\begin{equation}\label{eq:mix.boundary}
\intdPt (\bsvP - \bsvdP) \cdot \bsts \da_\tau \le 0, \qquad \intdPt (\dvphiP - \dvphidP) \mskip3mu \xis \da_\tau \le 0, \qquad \text{and} \qquad \intdPt (\mudP - \muP) \mskip3mu \bsjP \cdot \bs{n} \da_\tau \le 0.
\end{equation}

\subsection{Essential dynamic boundary conditions}

Essential boundary conditions result from the prescription of the surface fields onto the bulk fields. That is,
\begin{equation}\label{eq:bnd.cnd.essential}
\left.
\begin{aligned}
\bsvP = \bsvdP, \\[4pt]
\vphiP = \vphidP, \\[4pt]
\muP   = \mudP,
\end{aligned}
\right\} \quad \text{on } \dprt_\tau^{\mathrm{ess}}.
\end{equation}


\subsection{Natural dynamic boundary conditions}

Natural boundary conditions arise by specifying that the normal component of the bulk stress (described by the surface traction), the bulk microstress (described by the surface microtraction), and the bulk species flux are, respectively, equal to the normal component of the surface stress, the surface microstress, and the surface species flux on $\dprt^{\mathrm{nat}}_{\tau}$. In general, these are given by
\begin{equation}\label{eq:bnd.cnd.natural}
\left.
\begin{aligned}
\bsts &= \br{H}\bs{n}, \\[4pt]
\xis &= \bs{\tau}\cdot\bs{n}, \\[4pt]
\bsjP \cdot \bs{n} &= \bsjdP \cdot \bs{n},
\end{aligned}
\right\} \quad \text{on } \dprt_\tau^{\mathrm{nat}}.
\end{equation}
However, we have that $\br{H}\bs{n}=\bs{0}$, due to \eqref{eq:stresses.revised}$_2$ in addition to frame-indifference requirements. Furthermore, we have $\bs{\tau} \cdot \bs{n} =0$ and $\bsjdP \cdot \bs{n}=0$ as a result of adopting a tangential surface microstress $\bs{\tau}$ and a tangential surface mass flux $\bsjdP$. Thus, the natural boundary conditions read 
\begin{equation}\label{eq:bnd.cnd.natural.revised}
\left.
\begin{aligned}
\bsts &= \bs{0}, \\[4pt]
\xis  &= 0, \\[4pt]
\bsjP \cdot \bs{n} &= 0,
\end{aligned}
\right\} \quad \text{on } \dprt_\tau^{\mathrm{nat}},
\end{equation}
which satisfy the requirements in  \eqref{eq:free.energy.environment.mix}.

\subsection{Mixed dynamic boundary conditions}

In view of \eqref{eq:free.energy.environment.mix}, we formulate a set of mixed boundary conditions, which are dissipative in nature. These are given by
\begin{equation}\label{eq:bnd.cnd.mixed}
\left.
\begin{aligned}
\bsts &= \fr{1}{L_{\bs{\upsilon}}} \Prj{\bs{n}}(\bsvdP - \bsvP) = \fr{1}{L_{\bs{\upsilon}}} (\bsvdP - \bsvP), \\[4pt]
\xis &= \fr{1}{L_\varphi} (\dvphidP - \dvphiP), \\[4pt]
\bsjP \cdot \bs{n} &= - \fr{1}{L_\mu} (\mudP - \muP),
\end{aligned}
\right\} \quad \text{on } \dprt_\tau^{\mathrm{mix}},
\end{equation}
where $L_{\bs{\upsilon}}, L_\varphi, L_\mu > 0$. The expressions in \eqref{eq:bnd.cnd.mixed} should be understood as the surface traction, the surface microtraction, and the normal bulk species flux defined across $\dprt^{\mathrm{mix}}_{\tau}$, respectively, driven by the difference in velocity, microstructure (described by the phase field), and chemical potential between the bulk and surface material. Here, we assume that the dependency is linear and that the parameters $L_{\bs{\upsilon}}$, $L_\varphi$, $L_\mu$ act as relaxation parameters. Furthermore, note that the rightmost expression in \eqref{eq:bnd.cnd.mixed}$_1$ arises in view of assumption \ref{as:2}.

In \cite{Kno21b}, a mixed type of boundary condition similar to \eqref{eq:bnd.cnd.mixed}$_3$ was proposed. The theory for mixed dynamic boundary conditions for phase-field models was further extended and presented for both the microstructure and chemical potential in \cite{Esp23,espath2023mechanics}.

\section{Static edge boundary conditions}

We complement the dynamic boundary conditions on $\dprt_\tau$ previously presented by a set of static boundary conditions on $\ddprt_\tau$. In particular, we formulate the essential and natural boundary conditions resulting from the action of a static environment on the edge $\ddprt_\tau$, that is, the curve where the dynamic surface $\dprt_\tau$ loses smoothness. On static edges $\ddprt_\tau^{\mathrm{ess}}$, we first present the essential boundary conditions, which read
\begin{equation}\label{eq:edge.bnd.cnd.essential}
\left.
\begin{aligned}
\bsvdP  &=  \bsvddP^{\mathrm{env}}, \\[4pt]
\vphidP &=  \vphiddP^{\mathrm{env}}, \\[4pt]
\mudP   &=  \muddP^{\mathrm{env}},
\end{aligned}
\right\} \quad \text{on } \ddprt_\tau^{\mathrm{ess}},
\end{equation}
where $\bsvddP^{\mathrm{env}}$ is the action of the velocity, 
$\vphiddP^{\mathrm{env}}$ is the assignment of microstructure, and $\muddP^{\mathrm{env}}$ is the action of the chemical potential, all originating from a static environment acting on the edge $\ddprt_\tau^{\mathrm{ess}}$. 

Alternatively, as natural boundary conditions, we may prescribe
\begin{equation}\label{eq:edge.bnd.cnd.natural}
\left.
\begin{aligned}
\bshds \coloneqq \surp{\br{H} \bs{\nu}} = \bshds^{\mathrm{env}}, \\[4pt]
\tauds \coloneqq \surp{\bs{\tau} \cdot \bs{\nu}} = \tauds^{\mathrm{env}}, \\[4pt]
\jmath_{\scriptscriptstyle\ddprt} \coloneqq - \surp{\bs{\jmath} \cdot \bs{\nu}} = \jmath_{\scriptscriptstyle\ddprt}^{\mathrm{env}}.
\end{aligned}
\right\} \quad \text{on } \ddprt_\tau^{\mathrm{nat}},
\end{equation}
where $\bshds^{\mathrm{env}}$ and $\tauds^{\mathrm{env}}$ are the assigned edge traction and edge microtraction of the static environment across $\ddprt_{\mathrm{nat}}$. Furthermore, $\jmath_{\scriptscriptstyle\ddprt}^{\mathrm{env}}$ represents the transfer of species from the static environment to $\ddprt_\tau^{\mathrm{nat}}$.

\section{Dissipation inequalities}

In this section, we are interested in understanding the behavior of the total energy variation, that is,
\begin{equation}
\dot{\overline{\intPt\rhoP(\psiP+\fr{1}{2}|\bsvP|^2)\dv_\tau}} + \dot{\overline{\intdPt\rho\ell_\tau(\psidP+\fr{1}{2}|\bsvdP|^2)\da_\tau}}.
\end{equation}
To simplify and understand the decay of each energetic contribution, that is, the free energy and kinetic energy densities, we will spit these into
\begin{equation}\label{eq:energy.contrib}
\dot{\overline{\intPt\rhoP\psiP\dv_\tau}} + \dot{\overline{\intdPt\rho\ell_\tau\psidP\da_\tau}},  \qquad \text{and} \qquad
\dot{\overline{\intPt\fr{1}{2}\rhoP|\bsvP|^2\dv_\tau}} + \dot{\overline{\intdPt\fr{1}{2}\rho\ell_\tau|\bsvdP|^2\da_\tau}}.
\end{equation}

Emulating the procedure presented by Espath \cite{Esp23} for the temporal change in free-energy functional, we arrive at 
\begin{alignat}{2}\label{eq:free.energy.rate}
\dot{\overline{\intPt\rhoP\psiP\dv_\tau}} {}+{}&{} \dot{\overline{\intdPt\rho\ell_\tau\psidP\da_\tau}} {} \nonumber \\[4pt]
{}=& \intPt \left( \muP \sP + \gamma\dvphiP \right) \dv_\tau - \intPt \Grad \muP \cdot \bsMP \Grad \muP \dv_\tau - \intPt \left(\Grad\vphiP \otimes \bsxi \right) \twovdots \Grad \bsvP \dv_\tau \nonumber \\[4pt] 
&+ \intdPt \left(\mudP \sdP + \zeta \dvphidP \right) \da_\tau - \intdPt \Grads \mudP \cdot \bsMdP \Grads \mudP \da_\tau \nonumber \\[4pt] 
&- \intdPt \rho\ell_\tau{}^2 \psi~ \Divs \bsvdP  \da_\tau \nonumber - \intdPt \left(\Grads\vphidP \otimes \bs{\tau} \right) \twovdots \Grads \bsvdP \da_\tau 
\nonumber \\[4pt]
&- \intdPt \left(\left(\dvphidP - \dvphiP\right) \xis + (\muP - \mudP) \bsjP \cdot \bs{n} \right)\da_\tau   + \intddPt \left( \dvphidP \surp{\bs{\tau} \cdot \bs{\nu}} - \mudP \surp{ \bsjdP \cdot \bs{\nu}} \right)\ds_\tau.
\end{alignat}

Next, we consider the temporal change in kinetic energy \eqref{eq:energy.contrib}$_2$. For this purpose, we consider the decomposition into inertial and non-inertial parts \eqref{eq:generalised.body.force} and \eqref{eq:inertial.body.force}. Then, the variation in kinetic energy rate reads
\begin{alignat}{2}\label{eq:kinetic.energy.rate}
\dot{\overline{\intPt\tfrac{1}{2}\rhoP |\bsvP|^2_{}\dv_\tau}} +{}& \dot{\overline{\intdPt\tfrac{1}{2}\rho\ell_\tau|\bsvdP|^2_{}\da_\tau}} \nonumber\\[4pt]
={}& \intPt\rhoP\bsvP\cdot\dbsvP\dv_\tau + \intdPt\rho\ell_\tau\bsvdP\cdot\dbsvdP\da_\tau, \nonumber\\[4pt]
={}& \intPt\bsvP\cdot(\bsbni + \Div\brT)\dv_\tau + \intdPt\bsvdP\cdot(\bsgni - \bsts + \Divs(\brH\Prj{\bs{n}}))\da_\tau, \nonumber\\[4pt]
={}& \intPt(\bsvP\cdot\bsbni - \br{T}\twovdots\Grad\bsvP)\dv_\tau + \intdPt(\bsvdP\cdot\bsgni - \br{H}\Prj{\bs{n}}\twovdots\Grads\bsvdP )\da_\tau \nonumber\\[4pt]
&- \intdPt (\bsvdP -\bsvP)\cdot\bsts \da_\tau + \intddPt \bsvdP\cdot\surp{\brH\bs{\nu}} \ds_\tau,
\end{alignat}
where we have used the field equations \eqref{eq:field.equations.fluid.mod} and surface traction \eqref{eq:tractions}$_{1}$.

In view of the free- and kinetic-energy rates in \eqref{eq:free.energy.rate} and \eqref{eq:kinetic.energy.rate}, and by symmetry of $\brT$ and $\brH\Prj{n}$, the variation in total energy reads
\begin{align}\label{eq:total.energy.rate}
\dot{\overline{\intPt\rhoP(\psiP+\fr{1}{2}|\bsvP|^2)\dv_\tau}} +{}&{} \dot{\overline{\intdPt\rho\ell_\tau(\psidP+\fr{1}{2}|\bsvdP|^2)\da_\tau}}  \nonumber \\[4pt]
&\hspace{-50pt}= \intPt \left( \muP \sP + \gamma\dvphiP \right) \dv_\tau - \intPt \Grad \muP \cdot \bsMP \Grad \muP \dv_\tau - \intPt \left(\br{T} + \Grad\vphiP \otimes \bsxi \right) \twovdots \DP \dv_\tau \nonumber \\[4pt]
&\hspace{-38pt}+ \intPt \bsvP\cdot\bsbni \dv_\tau + \intdPt \left(\mudP \sdP + \zeta \dvphidP \right) \da_\tau - \intdPt \Grads \mudP \cdot \bsMdP \Grads \mudP \da_\tau \nonumber \\[4pt] 
&\hspace{-38pt}- \intdPt \rho\ell_\tau{}^2 \psi~ \Divs \bsvdP  \da_\tau \nonumber - \intdPt \left(\br{H}\Prj{\bs{n}} + \Grads\vphidP \otimes \bs{\tau} \right) \twovdots \DdP \da_\tau + \intdPt \bsvdP\cdot\bsgni \da_\tau
\nonumber \\[4pt]
&\hspace{-38pt} - \intdPt \left(\left(\dvphidP - \dvphiP\right) \xis + (\muP - \mudP) \bsjP \cdot \bs{n} + (\bsvdP -\bsvP)\cdot\bsts \right)\da_\tau   \nonumber \\[4pt]
&\hspace{-38pt}+ \intddPt \left( \dvphidP \surp{\bs{\tau} \cdot \bs{\nu}} - \mudP \surp{ \bsjdP \cdot \bs{\nu}} + \bsvdP\cdot\surp{\brH\bs{\nu}} \right)\ds_\tau,
\end{align}
where we recall that the stretching tensors are given by $\DP \coloneqq \sym \mskip2mu \Grad \bsvP$ and $\DdP \coloneqq \sym \mskip2mu \Grads \bsvdP$.
Substitution of the constitutive response for the bulk and surface stresses \eqref{eq:stresses.revised} and the thermodynamical surface pressure \eqref{eq.thermodynamical.pressure}, specializes \eqref{eq:total.energy.rate} to 
\begin{align}\label{eq:total.energy.rate.2}
\dot{\overline{\intPt\rhoP(\psiP+\fr{1}{2}|\bsvP|^2)\dv_\tau}} +{}&{} \dot{\overline{\intdPt\rho\ell_\tau(\psidP+\fr{1}{2}|\bsvdP|^2)\da_\tau}}  \nonumber \\[4pt]
&\hspace{-50pt}= \intPt \left( \muP \sP + \gamma\dvphiP \right) \dv_\tau - \intPt \Grad \muP \cdot \bsMP \Grad \muP \dv_\tau - \intPt 2 \bmudP |\DkP|^2 \dv_\tau \nonumber \\[4pt]
&\hspace{-38pt}+ \intPt \bsvP\cdot\bsbni \dv_\tau + \intdPt \left(\mudP \sdP + \zeta \dvphidP \right) \da_\tau - \intdPt \Grads \mudP \cdot \bsMdP \Grads \mudP \da_\tau \nonumber \\[4pt] 
&\hspace{-38pt}- \intdPt \left(2\bmudP |\Prj{\bs{n}}\DkdP|^2 + \barkappadP(\Divs \bsvdP)^2 \right) \da_\tau + \intdPt \bsvdP\cdot\bsgni \da_\tau
\nonumber \\[4pt]
&\hspace{-38pt} - \intdPt \left(\left(\dvphidP - \dvphiP\right) \xis + (\muP - \mudP) \bsjP \cdot \bs{n} + (\bsvdP -\bsvP)\cdot\bsts \right)\da_\tau   \nonumber \\[4pt]
&\hspace{-38pt}+ \intddPt \left( \dvphidP \surp{\bs{\tau} \cdot \bs{\nu}} - \mudP \surp{ \bsjdP \cdot \bs{\nu}} + \bsvdP\cdot\surp{\brH\bs{\nu}} \right)\ds_\tau,
\end{align}
where we have used the viscous dissipation identities \eqref{eq:viscous.dissipation}, the isochoric constraint $\tr\DP = \Div\bsvP = 0$ \eqref{eq:isochoric.constraint.bulk}, as well as the identity $\tr(\Prj{\bs{n}}\DdP) = \Divs \bsvdP$.

Lastly, we consider the set of mixed boundary conditions on $\dprt_\tau\coloneqq \dprt_\tau^{\mathrm{mix}}$ \eqref{eq:bnd.cnd.mixed} and natural boundary conditions on $\ddprt_\tau\coloneqq \ddprt_\tau^{\mathrm{nat}}$ \eqref{eq:edge.bnd.cnd.natural}. Then, in view of the residual dissipation inequality in \eqref{eq:res.diss.inequalities}, we arrive at the Lyapunov decay relation
\begin{align}\label{eq:total.energy.rate.incl.bnd.cnd}
\dot{\overline{\intPt\rhoP(\psiP+\fr{1}{2}|\bsvP|^2)\dv_\tau}} +{}&{} \dot{\overline{\intdPt\rho\ell_\tau(\psidP+\fr{1}{2}|\bsvdP|^2)\da_\tau}}  \nonumber \\[4pt]
&\hspace{-50pt}= \intPt \left( \muP \sP + \gamma\dvphiP \right) \dv_\tau - \intPt \Grad \muP \cdot \bsMP \Grad \muP \dv_\tau - \intPt 2 \bmudP |\DkP|^2 \dv_\tau \nonumber \\[4pt]
&\hspace{-38pt}+ \intPt \bsvP\cdot\bsbni \dv_\tau + \intdPt \left(\mudP \sdP + \zeta \dvphidP \right) \da_\tau - \intdPt \Grads \mudP \cdot \bsMdP \Grads \mudP \da_\tau \nonumber \\[4pt] 
&\hspace{-38pt}- \intdPt \left(2\bmudP |\Prj{\bs{n}}\DkdP|^2 + \barkappadP\Bigg(\dfrac{\mathring{\ell}_\tau}{\ell_\tau}\Bigg)^2 \right) \da_\tau + \intdPt \bsvdP\cdot\bsgni \da_\tau
\nonumber \\[4pt]
&\hspace{-38pt} - \intdPt \left(\fr{1}{L_\varphi} |\dvphidP - \dvphiP|^2 +\fr{1}{L_\mu} |\muP - \mudP|^2 + \fr{1}{L_{\bs{\upsilon}}}|\bsvdP -\bsvP|^2 \right)\da_\tau   \nonumber \\[4pt]
&\hspace{-38pt}+ \intddPt \left( \dvphidP  \tauds^{\mathrm{env}}  + \mudP \jmath_{\scriptscriptstyle\ddprt}^{\mathrm{env}} + \bsvdP\cdot\bshds^{\mathrm{env}} \right)\ds_\tau, \nonumber \\[4pt]
&\hspace{-50pt}\leq \intPt \left( \muP \sP + \gamma\dvphiP \right) \dv_\tau + \intPt \bsvP\cdot\bsbni \dv_\tau + \intdPt \left(\mudP \sdP + \zeta \dvphidP \right) \da_\tau + \intdPt \bsvdP\cdot\bsgni \da_\tau
\nonumber \\[4pt]
&\hspace{-38pt}+ \intddPt \left( \dvphidP  \tauds^{\mathrm{env}}  + \mudP \jmath_{\scriptscriptstyle\ddprt}^{\mathrm{env}} + \bsvdP\cdot\bshds^{\mathrm{env}} \right)\ds_\tau.
\end{align}

In \eqref{eq:total.energy.rate.incl.bnd.cnd}, we identify the following environmental contributions
\begin{itemize}
    \item $\muP \sP$ represents the rate at which energy is transferred to $\prt_\tau$ due to the production of species, and $\mudP \sdP$ represents the rate at which energy is transferred to $\dprt_\tau$ due to species production;
    \item $\gamma\dvphiP$ represents the power expended on the microstructure of $\prt_\tau$ by sources external to the body $\prt_\tau$, whereas $\zeta \dvphidP$ denotes the power expended 
    on the microstructure of $\dprt_\tau$ by sources external to the boundary of the body $\dprt_\tau$, which do not originate from $\prt_\tau$;
    \item $\bsvP\cdot\bsbni$ represents power expended in $\prt_\tau$ by the environment, whereas $\bsvdP\cdot\bsgni$ represents the power expended on $\dprt_\tau$ by the environment;
    \item 
    $\dvphidP \tauds^{\mathrm{env}}$ represents the power expended across $\ddprt_\tau$ by edge microtractions from the environment exterior to both $\dprt_\tau$ and $\prt_\tau$;
    \item 
    $\mudP \jmath_{\scriptscriptstyle\ddprt}^{\mathrm{env}} $ represents the energy exchange across $\ddprt_\tau$ induced by a tangent-normal species flux from the exterior of $\dprt_\tau$ and $\prt_\tau$;
    \item 
    $\bsvdP\cdot\bshds^{\mathrm{env}} $ represents the power expended across $\ddprt_\tau$ by edge tractions from the environment.
    \end{itemize}
Lastly, the following terms may contribute to dissipation in the bulk-surface system 
\begin{itemize}
    \item $ \Grad \muP \cdot \bsMP \Grad \muP $ and $\Grads \mudP \cdot \bsMdP \Grads \mudP$ represent dissipation due to species diffusion in $\prt_\tau$ and $\dprt_\tau$, respectively;
    \item the term $2\bmudP |\Prj{\bs{n}}\DkdP|^2$ represents the viscous dissipation in $\prt_\tau$, whereas $2\bmudP |\Prj{\bs{n}}\DkdP|^2 +$
    $\barkappadP(\mathring{\ell}_\tau / \ell_\tau)^2$ denotes the viscous dissipation in $\dprt_\tau$;
    \item $\tfrac{1}{L_\varphi}|\dvphidP - \dvphiP|^{2}_{}$ represents power expended across $\dprt_\tau$ driven by the difference in microstructure (described by the phase field) between $\dprt_\tau$ and the adjacent $\prt_\tau$;
    \item $-\tfrac{1}{L_\mu}|\mudP-\muP|^{2}_{}$ represents energy exchange across $\dprt_\tau$ due to the difference in chemical potential between $\dprt_\tau$ and the adjacent $\prt_\tau$;
    \item $\fr{1}{L_{\bs{\upsilon}}}|\bsvdP -\bsvP|^{2}_{}= \fr{1}{L_{\bs{\upsilon}}}|\Prj{\bs{n}}(\bsvdP -\bsvP)|^{2}_{}$ represents power expended across $\dprt_\tau$ driven by the difference in the tangential velocity components between $\dprt_\tau$ and the adjacent $\prt_\tau$.
\end{itemize}

\section{Acknowledgements}

This work was partially supported by the Flexible Interdisciplinary Research Collaboration Fund at the University of Nottingham Project ID 7466664. The research by KvdZ was supported by the Engineering and Physical Sciences Research Council (EPSRC), UK under Grant EP/W010011/1.

\section{Declarations: Funding and/or Conflicts of interests/Competing interests \& Data availability}

The authors have no conflicts to disclose.

Data sharing is not applicable to this article as no new data were created or analyzed in this study.

No AI tools were used in this work.

\section{Additional Declaration}

This research paper is a component of Anne Boschman's doctoral thesis. Interested readers seeking comprehensive understanding and additional contextual information are directed to \cite{Bos23}. The thesis offers a comprehensive exploration of the subject including supplementary insights that enrich the discussions presented in this paper.





\footnotesize
    

\begin{thebibliography}{10}

    \bibitem{Bou13}
    M~Boussinesq.
    \newblock Sur l'existence d'une viscosit\'e superficielle, dans la mince couche de transition s\'eparant un liquide d'une autre fluide contigu.
    \newblock {\em Ann. Chim. Phys.}, 29:349--357, 1913.
    
    \bibitem{Bot10}
    D~Bothe and J~Pr{\"u}ss.
    \newblock On the two-phase navier--stokes equations with boussinesq--scriven surface fluid.
    \newblock {\em Journal of Mathematical Fluid Mechanics}, 12(1):133--150, 2010.
    
    \bibitem{Lev62}
    V~Levich.
    \newblock {\em Physicochemical Hydrodynamics}.
    \newblock Prentice-Hall, Englewood Cliffs, New Jersey, 1962.
    
    \bibitem{Ada41}
    NK~Adam.
    \newblock Physics and chemistry of surfaces.
    \newblock 1941.
    
    \bibitem{Ada67}
    AW~Adamson, AP~Gast, et~al.
    \newblock {\em Physical chemistry of surfaces}, volume 150.
    \newblock Interscience publishers New York, 1967.
    
    \bibitem{Scr60}
    LE~Scriven.
    \newblock Dynamics of a fluid interface equation of motion for {N}ewtonian surface fluids.
    \newblock {\em Chemical Engineering Science}, 12(2):98--108, 1960.
    
    \bibitem{Gur75}
    ME~Gurtin and AI~Murdoch.
    \newblock A continuum theory of elastic material surfaces.
    \newblock {\em Archive for rational mechanics and analysis}, 57:291--323, 1975.
    
    \bibitem{Lad17}
    B~Ladoux and R~M{\`e}ge.
    \newblock Mechanobiology of collective cell behaviours.
    \newblock {\em Nature reviews Molecular cell biology}, 18(12):743--757, 2017.
    
    \bibitem{Esp21c}
    L~Espath.
    \newblock On the control volume arbitrariness in the navier--stokes equation.
    \newblock {\em Physics of Fluids}, 33(1), 2021.
    
    \bibitem{Bra15}
    C~Brangwynne, P~Tompa, and R~Pappu.
    \newblock Polymer physics of intracellular phase transitions.
    \newblock {\em Nature Physics}, 11(11):899--904, 2015.
    
    \bibitem{Shi17}
    Y~Shin and C~Brangwynne.
    \newblock Liquid phase condensation in cell physiology and disease.
    \newblock {\em Science}, 357(6357):eaaf4382, 2017.
    
    \bibitem{Mad15}
    A~Madzvamuse, A~Chung, and C~Venkataraman.
    \newblock Stability analysis and simulations of coupled bulk-surface reaction--diffusion systems.
    \newblock {\em Proceedings of the Royal Society A: Mathematical, Physical and Engineering Sciences}, 471(2175):20140546, 2015.
    
    \bibitem{Dud23}
    F~Duda, F~Forte~Neto, and E~Fried.
    \newblock Modelling of surface reactions and diffusion mediated by bulk diffusion.
    \newblock {\em Philosophical Transactions of the Royal Society A}, 381(2263):20220367, 2023.
    
    \bibitem{Esp23}
    L~Espath.
    \newblock A continuum framework for phase field with bulk-surface dynamics.
    \newblock {\em Partial Differential Equations and Applications}, 4(1):1, 2023.
    
    \bibitem{Cer05}
    P~Cermelli, E~Fried, and ME~Gurtin.
    \newblock Transport relations for surface integrals arising in the formulation of balance laws for evolving fluid interfaces.
    \newblock {\em Journal of Fluid Mechanics}, 544:339--351, 2005.
    
    \bibitem{Sil13}
    M~\v{S}ilhav\'y.
    \newblock A direct approach to nonlinear shells with application to surface-substrate interactions.
    \newblock 1(2):211--–232, August 2013.
    
    \bibitem{Tom23}
    G~Tomassetti.
    \newblock A direct, coordinate-free approach to the mechanics of thin shells.
    \newblock {\em arXiv preprint arXiv:2305.08884}, 2023.
    
    \bibitem{Gur82}
    Morton~E Gurtin.
    \newblock {\em An introduction to continuum mechanics}.
    \newblock Academic press, 1982.
    
    \bibitem{Gur10}
    ME~Gurtin, E~Fried, and L~Anand.
    \newblock {\em The Mechanics and Thermodynamics of Continua}.
    \newblock Cambridge University Press, April 2010.
    
    \bibitem{Esp20}
    L~Espath, VM~Calo, and E~Fried.
    \newblock Generalized {S}wift--{H}ohenberg and phase-field-crystal equations based on a second-gradient phase-field theory.
    \newblock {\em Meccanica}, 55(10):1853--1868, 2020.
    
    \bibitem{Esp21a}
    L~Espath and V~Calo.
    \newblock Phase-field gradient theory.
    \newblock {\em Zeitschrift f{\"u}r angewandte Mathematik und Physik}, 72(2):1--33, 2021.
    
    \bibitem{Fos16}
    R~Fosdick.
    \newblock A generalized continuum theory with internal corner and surface contact interactions.
    \newblock {\em Continuum Mechanics and Thermodynamics}, 28:275--292, 2016.
    
    \bibitem{Fri06}
    E~Fried and ME~Gurtin.
    \newblock Tractions, balances, and boundary conditions for nonsimple materials with application to liquid flow at small-length scales.
    \newblock {\em Archive for Rational Mechanics and Analysis}, 182(3):513--554, 2006.
    
    \bibitem{Kno21b}
    P~Knopf and A~Signori.
    \newblock On the nonlocal {C}ahn--{H}illiard equation with nonlocal dynamic boundary condition and boundary penalization.
    \newblock {\em Journal of Differential Equations}, 280:236--291, 2021.
    
    \bibitem{espath2023mechanics}
    L~Espath.
    \newblock {\em Mechanics and Geometry of Enriched Continua}.
    \newblock Springer Nature, 2023.
    
    \bibitem{Bos23}
    AM~Boschman.
    \newblock {\em Fundamental Theories for Bulk-Surface Interactions: Evolving Interfaces, Phase-Field Adhesion and Mixed-Dimensional Flows}.
    \newblock University of Nottingham, Nottingham, UK, 2023.
    
\end{thebibliography}

\end{document}